\documentclass[
aps,prd,nofootinbib,reprint,preprintnumbers,superscriptaddress
]{revtex4}
\UseRawInputEncoding
\usepackage{supertabular}
\usepackage{lipsum}
\usepackage{blindtext}
\usepackage{amsfonts}
\usepackage{tensor}
\usepackage{graphicx}
\usepackage{pict2e}
\usepackage{balance}
\usepackage{blindtext}
\usepackage{caption}
\usepackage{tcolorbox}
\usepackage{subfig}
\usepackage{amssymb, amsmath,environ}
\usepackage{soul}
\usepackage[normalem]{ulem}
\usepackage{color}
\usepackage[colorlinks=true,urlcolor=black,linkcolor=black,citecolor=red]{hyperref}


\newcommand{\bse}{\begin{subequations}}
	\newcommand{\ese}{\end{subequations}}
\newcommand{\be}{\begin{equation}}
\newcommand{\ee}{\end{equation}}

\NewEnviron{eqsplit}{%
	\begin{equation}\begin{split}
	\BODY
	\end{split}\end{equation}
}

\makeatletter
\newcommand*\bigcdot{\mathpalette\bigcdot@{.5}}
\newcommand*\bigcdot@[2]{\mathbin{\vcenter{\hbox{\scalebox{#2}{$\m@th#1\bullet$}}}}}
\makeatother

\newcommand{\bea}{\begin{eqnarray}}
\newcommand{\eea}{\end{eqnarray}}
\newcommand{\ba}{\begin{array}}
	\newcommand{\ea}{\end{array}}

\begin{document}
	\preprint{MITP-23-048}		
	
\title{Signature of Non-Minimal Scalar-Gravity Coupling with an Early Matter Domination on the Power Spectrum of Gravitational Waves} 
	\author{Amirsalar Nikandish}
	\email[]{a\_nikandish@sbu.ac.ir}
	\affiliation{Department of Physics, Shahid Beheshti University, 1983969411, Tehran, Iran}
	\author{Shiva Rostam Zadeh}
	\email[]{sh\_rostamzadeh@ipm.ir}
	\affiliation{School of Particles and Accelerators, Institute for Research in Fundamental Sciences (IPM),\\ P.O. Box 19395-5531, Tehran, Iran}
	\author{Reza Naderi}
	\email[]{r\_naderi8@yahoo.com}
	\affiliation{Department of Physics, Faculty of Basic Sciences,
		Azarbaijan Shahid Madani University, 53714-161, Tabriz, Iran}
	\author{Fatemeh Elahi}
	\email[]{felahi@uni-mainz.de}
	\affiliation{PRISMA$^+$ Cluster of Excellence \& Mainz Institute for Theoretical Physics,\\ Johannes Gutenberg University, 55099 Mainz, Germany}
	\author{Hadi Mehrabpour}
	\email[]{hadi.mehrabpour.hm@gmail.com}
	\affiliation{PRISMA$^+$ Cluster of Excellence \& Mainz Institute for Theoretical Physics,\\ Johannes Gutenberg University, 55099 Mainz, Germany}
	\affiliation{Center for High Energy Physics, Peking University, Beijing 100871, China}
	\affiliation{Department of Physics and State Key Laboratory of Nuclear Physics and Technology,\\ Peking University, Beijing 100871, China}
\begin{abstract}
	The signal strength of primordial gravitational waves experiencing an epoch of early scalar domination is reduced with respect to radiation domination. In this paper, we demonstrate that the specific pattern of this reduction is sensitive to the coupling between the dominant field and gravity. When this coupling is zero, the impact of early matter domination on gravitational waves is solely attributed to the alteration of the Hubble parameter and the scale factor. In the presence of non-zero couplings, on the other hand, the evolution of primordial gravitational waves is directly affected as well, resulting in a distinct step-like feature in the power spectrum of the gravitational wave as a function of frequency. This feature serves as a smoking gun signature of this model. In this paper, we provide an analytical expression of the power spectrum that illustrates the dependence of power spectrum on model parameters and initial conditions. Furthermore, we provide analytical relations that specify the frequency interval in which the step occurs. We compare the analytical estimates with numerical analysis and show they match well. 
\end{abstract}

\maketitle
\section{Introduction}
\indent Since the first detection of gravitational waves (GWs) at the Laser Interferometer Gravitational-Wave Observatory (LIGO) and Virgo, a new window to unravel the mysteries of the cosmos has opened \cite{abbott2016observation, harry2010advanced, aasi2015advanced, acernese2014advanced}. Even though the sources of the detected GWs have been astrophysical thus far \cite{abbott2019gwtc}, we hope to also detect the cosmological ones with the advance of the detectors \cite{maggiore2000gravitational}. Among the possible sources of GWs, the stochastic gravitational wave background (SGWB) originating from inflation is of particular interest, because detecting it may shed light on the history of the early universe\cite{grishchuk1975amplification, starobinskii1979spectrum, rubakov1982graviton,mcwilliams2019astro2020,guzzetti2016gravitational, caprini2018cosmological,riotto2002inflation,lino2022gravitational, baumann2009tasi,maggiore2018gravitational,caldwell2019astro2020, kalogera2019deeper,cornish2019discovery, shoemaker2019gravitational}.
The standard model of cosmology, under the assumption of radiation domination (RD) from inflation to matter-radiation equality, predicts an almost scale invariant power spectrum for the SGWB across all frequencies \cite{kawasaki2000mev,hannestad2004lowest,
%高橋史宜2005oscillation,
de2008new,de2015bounds}. 
Yet, many well-motivated proposals predict a transient phase, dominated by a different energy component, intervening between inflation and the Big Bang Nucleosynthesis (BBN) \cite{muia2023testing,d2019imprint}. Depending on the temperature range and other specific features of this transient period, the SGWB profile is expected to vary. 
Inspired by numerous extensions to the standard model of particle physics, we focus our attention on early matter domination \cite{peccei2008strong, kugo1984superpotential, banks2003supersymmetry, kawasaki2008solving, harigaya2013peccei, harigaya2015peccei, d2016supersymmetric, co2017saxion,coughlan1983cosmological, ellis1986axion,froggatt1979hierarchy, Elahi:2020pxl, Elahi:2021pug}. Specifically, we assume a scalar field, $\phi$, which behaves like a pressureless fluid scaling as $a^{-3}$ with $a$ being the scale factor, causes a transient matter domination era. Once $\phi$ decays, the universe reverts to the RD era again. Assuming a minimal coupling between $\phi$ and gravity, SGWB is only affected indirectly and because of the alteration of the scale factor and the Hubble rate \cite{muia2023testing, d2019imprint}. We coin the term \textit{Indirect effect} to denote this influence. In the context of early matter domination, the SGWB power spectrum is suppressed at high frequencies compared to standard cosmology, and the Equation of State (EoS) of the Universe governs the slope of the power spectrum \cite{d2019imprint}.

In this paper, we study early matter domination where the dominating field has a non-minimal coupling with gravity; i.e., $f(R,\phi)=\Big(\frac{1}{8\pi G}-\xi {\phi}^2\Big)R$. In the term $\xi R \phi^2$, $\xi$ signifies the gravitational coupling constant and $R$ denotes the Ricci scalar. This coupling stands as the sole feasible local, scalar interaction of its kind with the appropriate dimensions \cite{birrell1984quantum}. A coupling between a scalar field and gravity is proposed in numerous models aiming to resolve some of the inherent problems with inflation, reheating, and baryogenesis \cite{de2010f, capozziello2011extended, clifton2012modified, sotiriou2010f, jaime2012f, carroll2019spacetime, oikonomou2016f, odintsov2016gauss}. 
For instance, in the context of non-oscillatory inflationary models\cite{felder1999inflation, spokoiny1993deflationary, peebles1999quintessential, ellis2021non, de2021review}, the term $\xi R \phi^2$ has been recently used in an efficient reheating scenario denoted as Ricci reheating\cite{figueroa2017standard, dimopoulos2018non, opferkuch2019ricci, laverda2023ricci}. In the same context, a novel quintessential Affleck-Dine (AD) baryogenesis scenario has also been proposed\cite{bettoni2018quintessential} based on the term $\xi R \phi^2$, avoiding troublesome iso-curvature modes found in conventional AD scenarios.
 Moreover, the widely used self-interaction term for scalar fields, $\lambda \phi^4$, necessitates the inclusion of $\xi R \phi^2$ in the Lagrangian for proper renormalization in curved space-time \cite{birrell1984quantum, parker2009quantum}. Within the scalar sector of the Standard Model (SM) Lagrangian (Higgs) \cite{figueroa2017standard}, $\xi R \phi^2$ emerges as the missing term that upholds all the symmetries of both gravity and the SM.

As $\xi$ is subject to running, it cannot be universally set to zero across all energy scales \cite{figueroa2021dynamics}.  Two values of $\xi=0$ for minimal coupling and $\xi=1/6$ for conformal coupling are of particular interest. The latter, for the case $m_{\phi}=0$, leads to conformal invariance of the action and hence the field equation of motion\cite{birrell1984quantum, ford2021cosmological}. Nonetheless, no compelling reason supports the presence of minimal or conformal coupling in the real world, as no symmetry is enhanced \cite{carroll2019spacetime}. Furthermore, since $\xi$ is dimensionless, there is no reason for it to be small. It could be non-minimal, i.e. of the order of unity or more.\footnote{The
values of $\xi =10,~50$ and etc, are used in some recent studies\cite{dimopoulos2018non, opferkuch2019ricci, figueroa2021dynamics}.}\cite{carroll2019spacetime}. Current experiments place a weak constraint on $\xi$ ($\xi<10^{15}$) due to the feeble gravitational interaction with the SM fields\ \cite{figueroa2021dynamics, atkins2013bounds} However, GWs might offer insights into the existence and strength of such a coupling.

This paper delves into SGWB within a cosmological framework characterized by early matter domination, where the dominant field $\phi$ directly couples with gravity ($\xi R \phi^2$). D'Eramo et al. previously examined the case of $\xi =0$ \cite{d2019imprint}. To highlight distinctions from \cite{d2019imprint}, we specifically explore the non-minimal regime ($\xi \geq 1$). Studies indicate that such gravity couplings manifest as additional terms in the evolution equation of GWs \cite{hwang2001gauge, odintsov2022spectrum, de2010f, nishizawa2018generalized, arai2018generalized}. That is, alongside modifications to the scale factor and Hubble, i.e., \textit{indirect effect} \cite{bernal2020primordial}, an extra factor directly affects GW evolution. We term this extra impact the \textit{direct effect}. Employing Wentzel-Kramers-Brillouin (WKB) analysis, we elucidate the power spectrum resulting from GW propagation in our cosmological setting. Our numerical findings demonstrate that the presence of the $\xi R \phi^2$ term, with non-minimal coupling, deepens the kink shape in the power spectrum, compared with the case studied in \cite{d2019imprint}. Furthermore, owing to the \textit{direct effect}, an additional step-like feature emerges, resulting in an enhanced reduction in the power spectrum. We highlight several benchmarks that could be probed by upcoming gravitational wave experiments.

To deepen our understanding of the $\xi R \phi^2$ term,  it is crucial to provide an analytical interpretation of the resulting power spectrum. Hence, we use reasonable approximations to get an analytical expression for the power spectrum compared with the case of the standard cosmology and explain how each part of the spectrum in a specific frequency interval is shaped due to a specific physical effect that was dominant in the corresponding temperature interval. To this end, the high-frequency modes that became sub at high temperatures and thus got affected by the changes in the cosmological evolution are of particular interest. In particular,  we determine the fraction of the spectrum of high frequencies to that of low-frequency modes that did not experience any changes in the cosmological history. In this context, we utilize a physical quantity called the \textit{dilution factor}, which was introduced in \cite{d2019imprint}. While this analytical interpretation was performed for the case of $\xi = 0$ in \cite{d2019imprint}, exhibiting notable agreement with numerical results, our study expands this analysis to encompass the case of $\xi \neq 0$. The analytical formula we derive, expressed in terms of dilution and damping factors, closely aligns with numerical results, maintaining an acceptable level of accuracy.

Finally, we comment on the potential testability of this scenario, highlighting a selection of gravitational wave experiments that could probe these changes \cite{haque2021decoding}. These include the Laser Interferometer Space Antenna (LISA), which operates within the frequency range of $10^{-5}-1$ Hz \cite{amaro2017laser, amaro2012elisa, caldwell2019using}, the DECi-hertz Interferometer Gravitational wave Observatory (DECIGO) spanning $10^{-3}-10$ Hz \cite{sato2017status, seto2001possibility, kawamura2021current}, the Einstein Telescope (ET) operating within $1-10^4$ Hz \cite{punturo2010einstein, sathyaprakash2012scientific, hild2011sensitivity}, and the forthcoming Big Bang Observatory (BBO) encompassing $10^{-3}-10$ Hz\cite{crowder2005beyond, corbin2006detecting, smith2017sensitivity}. Other experiments will explore intermediate frequencies like square kilometer array or SKA, probing $10^{-9}-10^{-6} Hz$\cite{weltman2020fundamental, barausse2020prospects} and NANOGrav collaboration\cite{pol2021astrophysics, arzoumanian2020nanograv} that works based on pulsar timing arrays (PTA) measurements \cite{odintsov2022spectrum, haque2021decoding}.\footnote{Recently, the evidence for GWs background in nano$Hz$ frequency is presented by the PTA data releases\cite{lee2023searching, antoniadis2023second, agazie2023nanograv, reardon2023search}.}

This paper is organized as follows: In Sec.\ \ref{NNSC}, we introduce the model and the cosmological framework in detail. In Sec.\ \ref{sec:GWeq}, we study the evolution of GWs, highlighting the effect of $\xi R \phi^2$.  We present the WKB solution for the modes that become sub, in Sec.~\ref{EGW}. We follow in Sec.~\ref{subsecPS}, by introducing the power spectrum as the observable of GWs, and in Sec.~\ref{sec33}, we highlight two of the pivotal frequencies in the power spectrum. Sec.~\ref{results} is dedicated to numerical results for the power spectrum and a comparison with the analytical estimates. Finally, the concluding remarks are presented in Sec.~\ref{con}.

\section{Non-minimally coupled scalar field }\label{NNSC}

Our theory is defined by a real scalar field $\phi$ that has feeble interactions with other fields, but has a non-minimal coupling with gravity: $f(R,\phi)=\Big(\frac{1}{8\pi G}-\xi {\phi}^2\Big)R$. The action of our theory is, thus, \cite{opferkuch2019ricci, figueroa2021dynamics}:\footnote{The most general $f(R,\phi)$ gravity action with an arbitrary number of scalar fields generally coupled with gravity has been discussed in Ref.~\cite{hwang2001gauge}. Note that authors consider the theory was included a non-linear sigma-type kinetic term which is different from what we consider here.}
\begin{align}\label{actionsp}
& \mathcal{S}=\int d^4 x \sqrt{-g}\left(\frac{1}{16\pi G}R-\frac{1}{2}g^{\mu\nu}\partial_\mu \phi \partial_\nu \phi -\frac{1}{2}\xi R \phi^2 -\frac{1}{2}m^2_\phi \phi^2+\mathcal{L}_M\right),
\end{align}
where $g$ is the determinant of the metric $g_{\mu \nu}$, $G$ is the gravitational coupling constant, $m_\phi$ is the mass of $\phi$, and $\mathcal{L}_M$ contains the kinetic component and the interactions of any other field in the cosmos, including its interaction with the scalar field, $\phi$. Varying Eq.\ (\ref{actionsp}) with respect to $g_{\mu \nu}$, one can obtain the gravitational equation as \cite{figueroa2021dynamics, hwang2001gauge}:
\begin{align}\label{E2}
\begin{split}
G_{\mu\nu} &= 8 \pi G {T^{(\text{eff})}_{\mu\nu}} = 8 \pi G \left({T^{(M)}_{\mu\nu}} + {T^{(\phi)}_{\mu\nu}}\right).
\end{split}
\end{align}
In the definition of Eq.\ (\ref{E2}), the energy-momentum tensor of the scalar filed $\phi$ is:
\begin{equation}\label{2-12}
T_{\mu\nu}^{(\phi)}=\partial_\mu \phi \partial_\nu \phi -g_{\mu\nu}\left(\frac{1}{2}\partial^\gamma \phi \partial_\gamma \phi +\frac{1}{2}m_\phi^2 \phi^2 \right)+\xi\left(G_{\mu\nu}+g_{\mu\nu}\Box-\nabla_\mu \nabla_\nu\right)\phi^2,
\end{equation}
where we have $\Box {\phi}^2=\frac{1}{\sqrt{-g}}\partial_\mu(\sqrt{-g}\partial^\mu {\phi}^2)$. Varying Eq.\ (\ref{actionsp}) with respect to $\phi$ yields the scalar field equation of motion \cite{hwang2001gauge}:
\begin{equation}\label{eq:phiEOM}
\Box \phi+(\xi R + m_\phi^2) \phi=-\frac{\partial \mathcal{L}_M}{\partial\phi}.
\end{equation}
Using Eqs. (\ref{E2}, \ref{eq:phiEOM}) and the Bianchi identity, $ \nabla_{\mu} G^{\mu}_{\ \nu} = \nabla_{\mu} T^{\mu (eff)}_{\ \nu}=0$, the continuity equation of the matter part can be obtained, 
$\nabla_\mu T_{\ \nu}^{\mu (M)}=\frac{\partial \mathcal{L}_M}{\partial \phi} \partial_\nu \phi$ \cite{hwang2001gauge}.
In these equations, the field $\phi$ is only a function of time in order to respect the homogeneity and isotropy of the Universe, and $\mathcal{L}_M$ describes the radiation. Therefore, the evolution equations for the scalar field $\phi$ and the energy density of radiation $\rho_R$ are:
\begin{align}
\ddot \phi + (3 H + \Gamma) \dot \phi + (\xi R + m_\phi^2) \phi = 0,\label{2.6}\\
\dot \rho_R + 4 \frac{g_\star^s}{g_\star^\rho}H \rho_R = \Gamma \dot \phi^2,\label{2.21}
\end{align}
where to obtain these equations, we have employed the flat Friedmann-Lemaitre-Robertson-Walker (FLRW) metric, $ds^2 = - dt^2 + a(t)^2 \delta_{ij} dx^i dx^j$,\footnote{In this study, our metric signature is space-positive $(-,+,+,+)$ and the equations are written in natural unit, where $\hbar = c = k_B = 1$.} with $i,j = 1-3$ specifying the spatial coordinates, explains the metric of a homogeneous and isotropic Universe, $a$ is the cosmic scale factor, and accordingly the Ricci scalar is obtained as $R= 6 (\dot a^2 + a \ddot a)/ a^2$ \cite{hwang2001gauge}.
Furthermore, $\Gamma$ represents the total decay rate of the scalar. In this paper, we are oblivious to the exact nature of the fields into which $\phi$ decays, and rather we assume it is part of radiation. The total number of relativistic degrees of freedom contributing to the energy (entropy) density of radiation is represented by $g_\star^\rho$ ($g_\star^s$). The evolution of the Hubble rate, defined as $H\equiv \frac{\dot a}{a}$, is described by the first Friedmann equation, obtained from Eq.\ (\ref{E2}) as: 
\begin{equation}\label{2.15}
H^2 = \frac {1}{3 M_{Pl}^{2}} \rho^{tot}=\frac{1}{3M_{Pl}^2}(\rho_\phi +\rho_R),
\end{equation}
where $M_{Pl}$ is the reduced Planck mass, and the energy density of $\phi$ can be obtained from the $T_{00}^{(\phi)}$ component of Eq.\ (\ref{2-12}) \cite{figueroa2021dynamics, opferkuch2019ricci}: 
\begin{equation}\label{eq:rhophi}
\rho_{\phi}=\frac 1 2 \dot{\phi}^2+\frac 1  2 m^2_\phi \phi^2+\xi \left(3 H^2 \phi^2+6 H {\phi} \dot{\phi}\right).\\
\end{equation}
Using 
%Eqs.\ (\ref{2.6}),(\ref{2.21}), and (\ref{2.15})
Eqs.\ (\ref{2.6}), (\ref{2.21}), and (\ref{2.15}), the system of coupled differential equations for the background would be complete for three unknowns, $\phi,\ \rho_R,$ and $H$. The free parameters in these equations are $$T_{in}, \hspace{0.2 in} \phi_{in} \equiv \phi(T_{in}),\hspace{0.2 in}  \dot \phi_{in} \equiv \dot \phi(T_{in}), \hspace{0.2 in}\xi, \hspace{0.2 in} m_
\phi, \hspace{0.2 in} \Gamma.$$ 
In this project, we fix $T_{in} = 10^{11} \ \text{GeV}$, motivated by well-studied reheating scenarios which prefer $T_{Reh} \in (10^{12} - 10^{15})\  \text{GeV}$\ \cite{kofman1997towards, felder1999instant, amin2015nonperturbative, lozanov2019lectures, figueroa2017standard, dimopoulos2018non}. 
The initial value of $\phi_{in}$ has an upper bound of $M_{Pl}$ to avoid undergoing the Universe to another inflationary period \cite{d2019imprint}. In addition, since the most noticeable deviations from standard cosmology occur when $
\phi_{in} \sim M_{Pl}$, we choose $\phi_{in} = 10^{18}$. 
The value of $\xi$, also, has an upper bound to make sure the Hubble rate stays real, i.e., $\xi<(M_{pl}^2/\phi_{in}^2)$, which in our case gives rise to $\xi_{max}\simeq 5.95$.\footnote{It is worth noting that in cases where $\phi_{in}$ is small, this constraint is not a limitation.}
Since our setup is not sensitive to $\dot \phi_{in}$ and any initial condition satisfying
\begin{equation}
\frac{1}{2} \dot \phi_{in}^2, \  \frac{1}{2} m_\phi^2 \phi_{in}^2,\frac{ 3 \xi^2 \dot \phi_{in}^2 \phi_{in}^2}{M_{Pl}^2 - \xi \phi_{in}^2 } \ll \rho_{R} (T_{in}),
\label{eq:inicond}
\end{equation}
leads to the total equation of state $ \omega \simeq 1/3$; and hence, a negligible $\dot \phi$ (see Appendix \ref{app:attractor}), we fix $\dot \phi_{in} = 0$ for simplicity.  
Concentrating on one of the benchmarks of Ref.~\cite{d2019imprint}, i.e. $m_\phi = 10$GeV and $\Gamma = 10^{-8}$GeV, we numerically solve the set of coupled equations and compare the results for the two cases of $\xi = 0$ (minimal coupling) \cite{d2019imprint} and $\xi \geq 1$ (non-minimal coupling).\footnote{It is due to the reasons described in the introduction.}

\subsection{General Evolution of the Background}\label{GEtB}

\begin{figure}
	\subfloat[]{\includegraphics[width=0.5\textwidth]{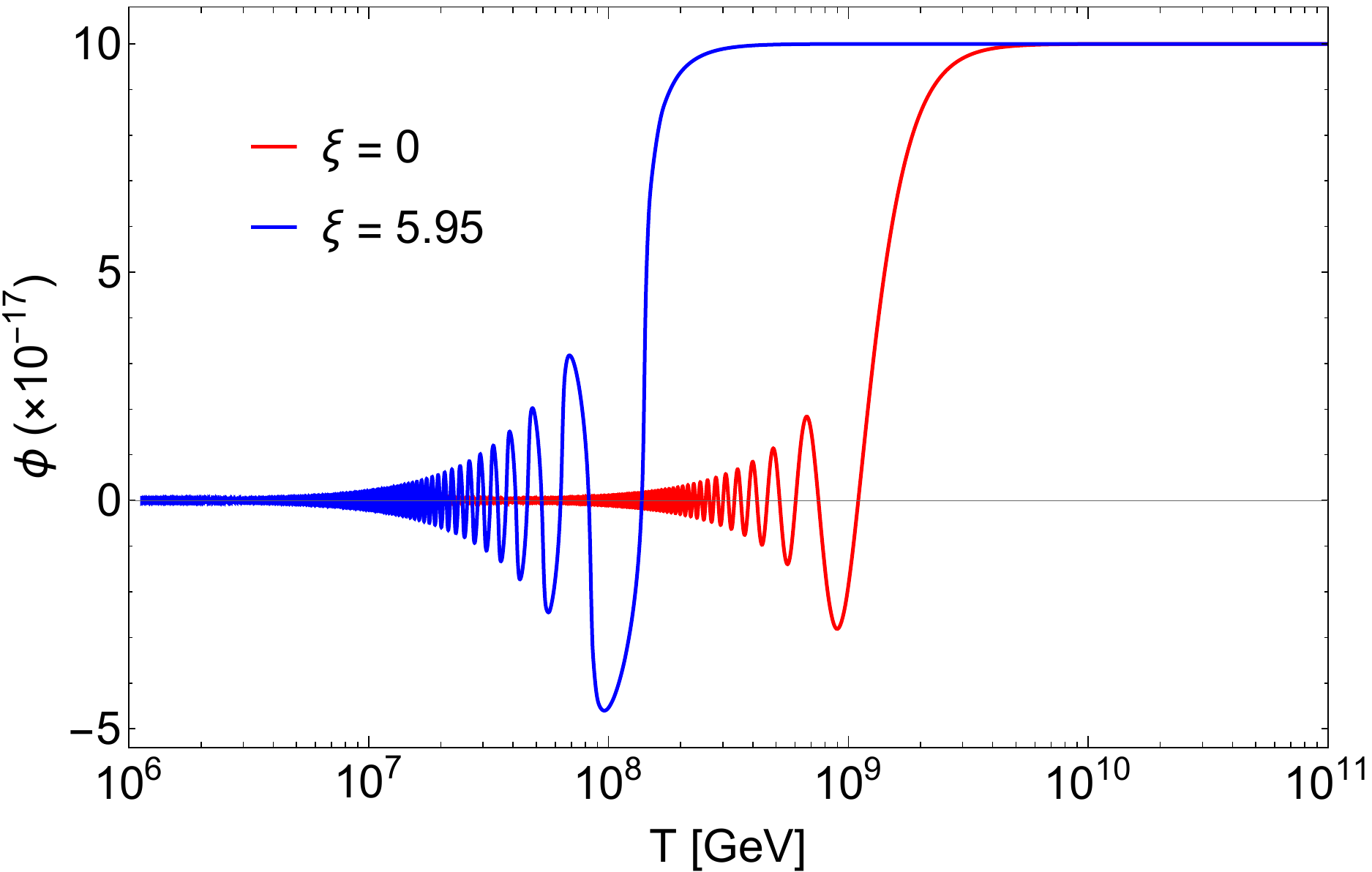}}
	\hfill
	\subfloat[]{\includegraphics[width=0.5\textwidth]{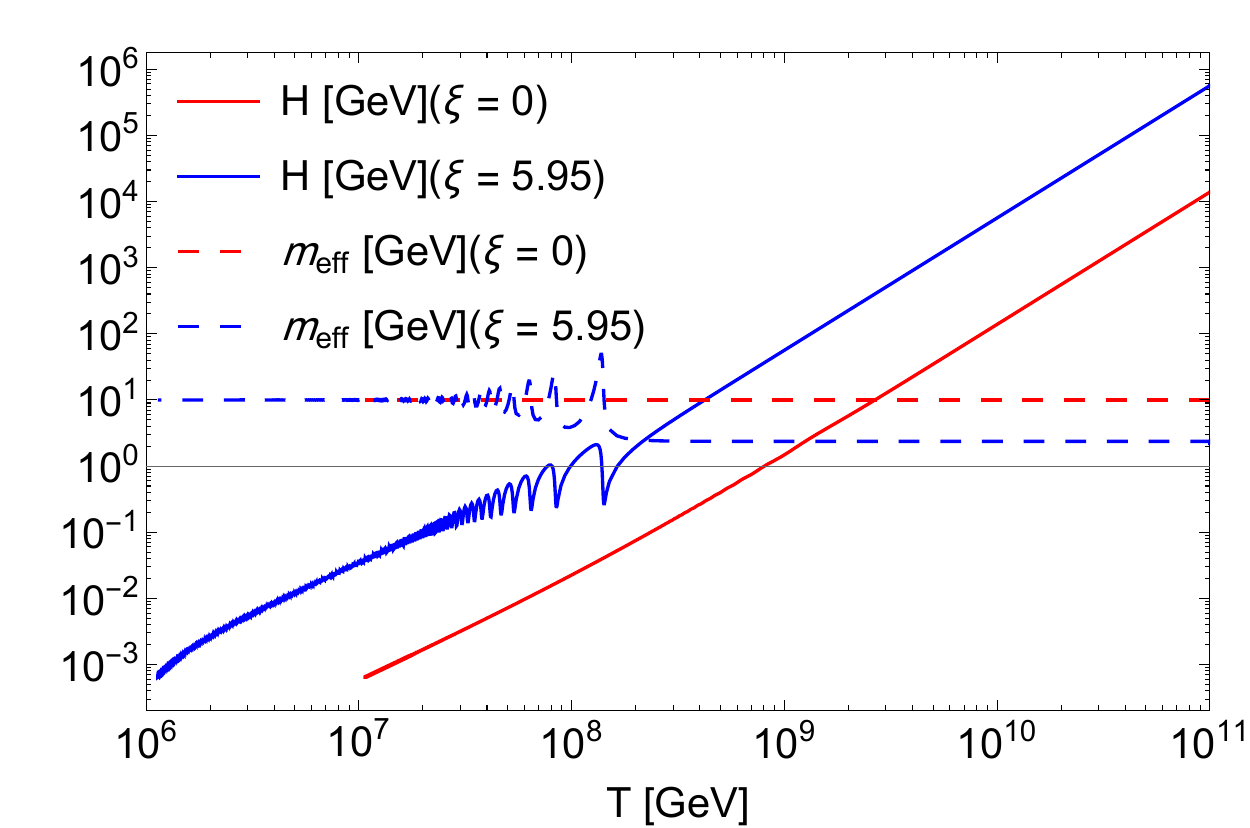}}
	\caption{(a) The behavior of the scalar field, $\phi$, in terms of temperature, $T$. (b) The evolution of the Hubble parameter and the effective mass as a function of temperature. In both plots, the red line is for $\xi = 0$, and the blue line is for $\xi = 5.95$. As shown in these plots, the Hubble rate increases and $m_{eff}$ decreases with the value of $\xi$, and therefore the temperature at which $ H \simeq m_{eff}$ is lower for larger values of $\xi$. Thereby, the oscillation of $\phi$ is delayed with non-minimal coupling of $\phi$ with gravity.}\label{fig:1}
\end{figure}

Given our initial conditions at high temperatures, the friction term in Eq.\ \ref{2.6} ( more specifically the Hubble term, $H \dot \phi$) forces $\phi$ to be stuck at its initial value,
similar to the case of $\xi = 0$ \cite{d2019imprint}. As the Hubble rate decreases, eventually the friction term becomes less efficient and $\phi$ starts oscillating when $H\simeq m_{eff} \equiv\sqrt{\xi R + m_\phi^2}$. As demonstrated in Fig.~\ref{fig:1}, the effect of gravitational coupling causes $\phi$ to remain close to its initial value for a longer duration,  because the Hubble rate is higher and $m_{eff}$, because of negative $R$, is lower for larger $\xi$:  $H\simeq \sqrt{\rho_R (T)(3 (M_{Pl}^2- \xi \phi^2_{in}))^{-1}}$ (see Appendix \ref{app:attractor} for this simplified expression of Hubble rate at high temperatures). 
 Around the time when $\phi$ starts oscillating, the universe deviates from state of RD, or equivalently   $\omega \simeq 1/3$, or equivalently the state of Radiation Domination (RD). The temperatures at which this occurs is 
\begin{equation}
T_{osc} \simeq \left(\frac{90}{i_{osc} \pi^2g^{\rho}_{*osc}}\right)^{1/4}(m_{\phi}M_{Pl})^{1/2}\left(\frac{1-\xi^2(\frac{\phi_{in}}{M_{Pl}})^4}{1+\xi(6\xi-1)(\frac{\phi_{in}}{M_{Pl}})^2}\right)^{1/4},
\label{eq:Tosc}
\end{equation}
when $H$ and $m_{eff}$ lines meet ($H\simeq m_{eff}$). 
The numerically determined factor $i_{osc} \simeq 6$ has been implemented to obtain a more precise value for $T_{osc}$, since $\phi$ oscillations start slightly after $H\simeq m_{eff}$ (e.g., for $\xi=0$, oscillations start at $T\sim10^9$ and for $\xi=5.95$, they start at $T\sim10^8$). After some oscillations,\footnote{After $\phi$ becomes oscillatory, solving the aforementioned coupled equations becomes rapidly more CPU-consuming. Thus, at $t \simeq 10^5/m_\phi$ when $\dot \phi \simeq m_\phi \phi$ and $\rho_\phi \simeq \dot \phi^2,$ we replace Eqs.\ (\ref{2.6}) and (\ref{2.21}) with the following approximate Boltzmann Equations:
\begin{align}\label{eq:evoleqapprox}
&\dot \rho_\phi + 3 H \rho_\phi   = - \Gamma \rho_\phi,\nonumber \\ 
&\dot \rho_R  + 4   \frac{g_{\star}^s }{g_{\star}^\rho} H \rho_R = \Gamma  \rho_\phi.  
\end{align}
Since $g_{\star}^s$ and $g_{\star}^\rho$ vary with temperature, especially at $T_{EW}\simeq 100GeV \leq T \leq T_{BBN}$, we evaluate them as a function of $\rho_R$ using Ref. \cite{saikawa2018primordial}.}  $\phi$ behaves as $ \frac{\sin (m_\phi t)}{m_\phi t}$. Hence, the energy density of $\phi$ redshifts like cold matter, leading to an era of matter domination (\textit{MD era}). This process continues until the decay of $\phi$ becomes more efficient than the Hubble rate, at which point $\phi$ depletes. We refer to this era, the \textit{ decay era} (DE). When $\rho_{\phi}$ becomes negligible with respect to $\rho_{R}$, we return to radiation domination, which we will refer to as the \textit{Late RD era}. The temperature at which we return to late RD era can be found by setting  $\rho_{\phi} \thicksim \rho_R$ \ \cite{d2019imprint}:
\begin{equation}
T_{RD} \simeq (\frac{\alpha_{RD}}{c_{RD}})^{1/2} \left(\frac{90}{2\pi^2g^{\rho}_{*RD}}\right)^{1/4}(\Gamma M_{Pl})^{1/2},
\label{eq:TRD}
\end{equation}
where $\alpha_{RD} \approx 0.64$.\footnote{Eq.\ \ref{eq:TRD} is derived assuming a constant EoS, or equivalently assuming $a \propto t^\alpha$, with $\alpha$ being a constant.} Assuming a constant $\omega$ throughout the MD era, $c_{RD}$ can be numerically determined as $c_{RD} \simeq 1.07$. At this point, the evolution of Universe is followed by the standard model of cosmology.

To investigate the differences between minimal ($\xi=0$) and non-minimal ($\xi \geq 1$) $\phi-$gravity coupling, GWs are the best candidates. The imprint of $\xi =0 $ case on GWs has been investigated in Ref.~\cite{d2019imprint}. Due to the non-minimal coupling between $\phi$ and gravity, the evolution of $\phi$ significantly impacts in the evolution of GWs, which will be discussed in details in the following section.

\section{The evolution of GWs and the power spectrum}\label{sec:GWeq}

Gravitational waves, $h_{ij}$, are perturbations in the space part of the FLRW metric \cite{riotto2002inflation, baumann2009tasi, gorbunov2011introduction, maggiore2018gravitational}:
\begin{equation}\label{2-3-1}
ds^2 = -dt^2 + a(t)^2 ({\delta}_{ij}^{(3)} + h_{ij}(t,\textbf{x})) dx^i dx^j,
\end{equation}
with $|h_{ij}| \ll 1$. Taking $h_{ij}$ to be transverse and traceless $\left( h^{TT}_{ii} = k_i h^{TT}_{ij} = 0\right)$, there are two degrees of freedom for $h_{ij}$ indicating two polarizations, $\zeta = +,\times$. By linearizing the gravitational equation, Eq.\ (\ref{E2}), the evolution equations of GWs is obtained:
\begin{equation}\label{2-3-11}
\frac{\partial^2 h^{TT}_{ij}}{\partial t^2} - \frac{1}{a^2} \nabla^2 h^{TT}_{ij} + \left(3H + A\right) \frac{\partial h^{TT}_{ij}}{\partial t} = 0,
\end{equation}
where $A = -2\xi \phi \dot{\phi}\big(\frac{1}{8\pi G} - \xi \phi^2\big)^{-1}$. It can be seen that an additional term, $A$, appears in the well-known evolution equations of the GWs \cite{riotto2002inflation,lino2022gravitational, baumann2009tasi,maggiore2018gravitational}, which comes from the $\xi R {\phi}^2$ term. It is through this term that the evolution of $\phi$ directly impacts the evolution of the SGWB.

It is conventional to expand $h_{ij}^{TT}$ as ${h_{ij}^{TT}}(t,x) = \sum_{\zeta} e_{ij}^{(\zeta)}{h^{(\zeta)}(t,x)}$ where $e_{ij}^{(\zeta)}$ is the polarization tensor\cite{gorbunov2011introduction, maggiore2018gravitational, riotto2002inflation}.\footnote{In our convention, $e_{ij}^{(+)} = \frac{1}{\sqrt{2}}(e_i^{(1)} e_j^{(1)} - e_i^{(2)}e_j^{(2)})$ and $e_{ij}^{(\times)} = \frac{1}{\sqrt{2}} (e_i^{(1)} e_j^{(2)} + e_i^{(2)} e_j^{(1)})$\cite{maggiore2018gravitational}. } Substituting the polarization expansion and ${h^{(\zeta)}}(t,x) = \int \frac{\text{d}^3 k}{(2\pi)^3} e^{ik\cdot x} {h^{(\zeta)}_k}(t)$ in Eq.\ (\ref{2-3-11}), the evolution equation for the amplitude of GWs, $h_k$ is obtained as following \cite{hwang2001gauge}:\footnote{Eq.\ (\ref{2.35}) is the same for both polarizations. Therefore, the index $\zeta$ is omitted in what follows.} 
\begin{equation}\label{2.35}
\ddot h_k+\left(3 H+A\right) \dot h_k+\frac{k^{2}}{a^{2}} h_k=0,
\end{equation}
where $h_k(t)$ represents the Fourier transform of $h(x,t)$.
For a better understanding of the behavior of GWs, Eq.\ (\ref{2.35}) is expressed in terms of conformal time,$\eta$, given by $\eta = \int \frac{dt}{a}$, as in Refs. \cite{odintsov2022spectrum, nishizawa2018generalized, arai2018generalized}:
\begin{equation}\label{2-3-47}
h_k^{\prime\prime}(\eta) + a(2H+A)h_k^{\prime}(\eta) + k^2 h_k(\eta) = 0,
\end{equation}
with the prime denoting differentiation with respect to conformal time. In what follows we do not mention the subscript k, since the term $k^2$ can highlight that the equation is written for a specific momentum.
The above equation is a harmonic oscillator with the damping term $\frac{a}{2} \left(2H + A\right)$. Using the numerical results of the background as obtained in Sec.~\ref{NNSC},\footnote{Due to the fact that GWs are just perturbations on the background and do not have any back-reaction on it, it is justified to solve the evolution equations of background separately and then the resulting $a$, $H$, and $A$ are inserted in the equation of GWs. In addition, this approach has two technical advantages: the background is solved only once, which significantly reduces the running time of the code, and fixes the precision of the numerical solution of the background for all modes.} we distinguish different regimes in the behavior of the damping term, by comparing the evolution of the terms $aA$ and $2aH$ as depicted in Fig.~\ref{fig:11}:

\begin{figure}
	\centering
	\includegraphics[width=.65\textwidth]{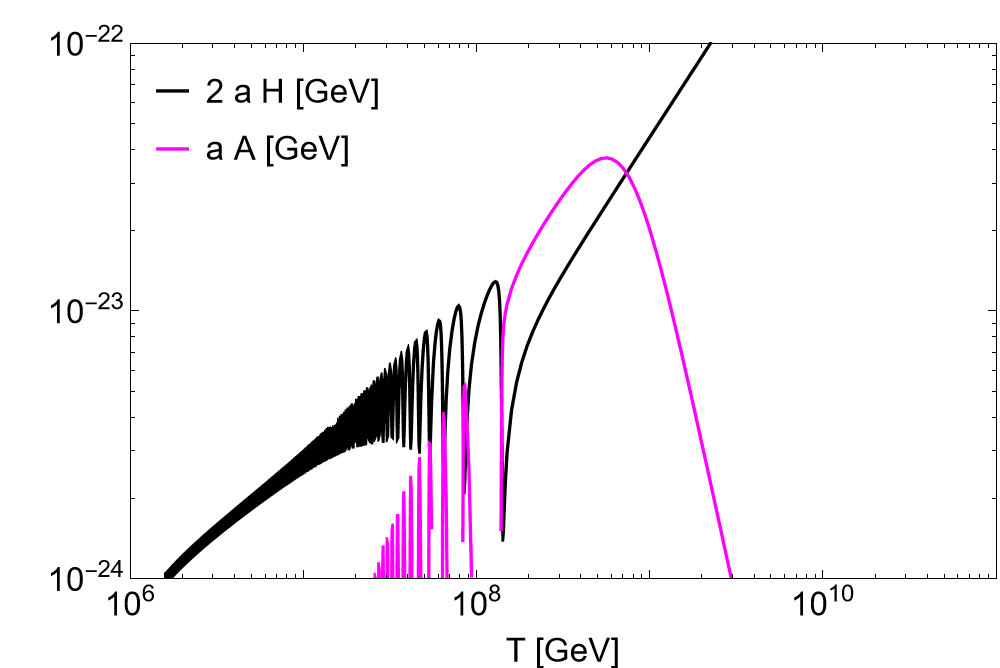}
	\caption{This plot illustrates the rates of $ 2aH$ and $aA$, which represent the damping factors in the evolution of GWs, for the case of $\xi = 5.95$. At high temperatures, $A \propto \dot \phi $ is zero. However, as $\phi$ gains momentum, $A$ rapidly grows and surpasses $H$ within certain temperature intervals. In addition, the ripples of $2aH$ and $aA$ become comparable at the onset of the oscillation era, and thus cancel each other out.}\label{fig:11}
\end{figure}

\begin{itemize}
	\item \textbf{$A \gtrsim 2H$ :} Even though $A$ is initially zero ($A \propto \dot \phi$ and $\dot \phi_{in} = 0$), it rapidly grows as $\phi$ gains momentum. Depending on the value of $\xi$ and our initial conditions, $A$ may eventually exceed the Hubble rate. We refer to this era as \textit{$A$ Domination era} (AD era) for short.\footnote{ Even though this Period effects the evolution of GWs, it does not change the EoS of the background. At the background level, the EoS parameter is still around $\frac{1}{3}$ which denotes that we are still in RD era.}
	During AD era, the evolution of GWs is not only affected by the altered background but also directly by $A$. To highlight the effect of $A$, we refer to its contribution as the \textit{direct effect}. Note that the direct effect goes to zero as $\xi \to 0$. Conversely, for large values of $\xi$  $A$ remains comparable to $H$ ($A \thicksim H$) during the first few oscillations, and strongly damps the oscillations.
	
	\item \textbf{$2H \gg A \neq 0$ :} The effect of $A$ rapidly diminishes, and the Hubble rate dominates the the damping term. In this case, the evolution of GWs is mainly influenced by the modified background. Since it is the effect of $\xi R {\phi}^2$ coupling that indirectly affects the evolution of GWs through $a$ and $H$, we denote this effect as \textit{indirect effect}.
	
	\item \textbf{$A\to 0$ :} This is the case in the late RD era, when $\phi$ is almost gone. Therefore, not only $A$ diminishes but also the evolution of background becomes that of the standard cosmology.
\end{itemize}
Different regimes leave different signatures on the GWs power spectrum, as we will show in Sec.~\ref{subsecPS}. For this purpose, let us first describe the behaviour of GWs at different frequency regimes.

\subsection{Evolution of GWs}\label{EGW}

Recalling the well-known theory of damping harmonic oscillators, the general behavior of a specific mode in Eq. (\ref{2-3-47}) can be understood by comparing its momentum, $k$, with the damping term, $\frac{a}{2}(2H+A)$. In this regard, we categorize different modes as:

\begin{itemize}
	\item \textbf{Sup Modes:} For a mode with $k \ll \frac{a}{2}(2H+A)$, there exists a constant solution to Eq.(\ref{2-3-47}), $h^{\text{Prim}}_k$, where the superscript ``Prim" stands for primordial. This solution describes the frozen behavior of the mode from the time it leaves the horizon during the inflationary epoch until it reaches the time $\eta_{\times}$ when $k \thicksim \frac{a_{\times}}{2}(2H_{\times}+A_{\times})$. The solution $h^{\text{Prim}}_k$ serves as an initial value for the subsequent evolution of the mode after $k \gg \frac{a}{2}(2H+A)$.
	
	\item \textbf{Sub Modes:} For a mode with $k \gg \frac{a}{2}(2H+A)$, it is conventional to parametrize the solution of Eq. (\ref{2-3-47}) as	$h_k (\eta) = h^{\text{Prim}}_k \chi_k(\eta)$, where $\chi_k(\eta)$ represents the transfer function that describes the subsequent evolution of the mode from $\eta_{\times}$ to the present time. By substituting this parametrization into Eq.(\ref{2-3-47}), the evolution equation for the transfer function becomes:
	\begin{equation}\label{transeq}
	\chi^{\prime\prime} + a (2H + A) \chi^\prime + k^2 \chi = 0,
	\end{equation}
	which has a general solution presented in Appendix \ref{WKBAp}. By imposing the initial conditions $\chi = 1$ and $\chi^{\prime} = 0$ for $k \ll \frac{a}{2}(2H+A)$, the specific oscillatory solution for Eq. (\ref{transeq}) can be expressed as:
	\begin{equation}\label{WKB}
	\chi_k(\eta) = e^{-d_k(\eta)} \chi_k^{GR}(\eta),
	\end{equation}
	where 
	\begin{align}\label{WKB2} 
	\chi_k^{GR}(\eta) = \frac{a_{\times}}{a(\eta)}e^{\pm ik\eta} \quad \textrm{and} \quad d_k(\eta) = \frac{1}{2}\int_{{\eta}_{\times}}^{\eta} aA\,d{\eta^{\prime}} = \frac{1}{2} \int_{{t}_{\times}}^{t} A\,dt^{\prime},
	\end{align}
\end{itemize}   
with ${d_k}(\eta)$ denoting the damping factor for a specific mode $k$ computed up to $\eta$. To calculate the observable, power spectrum of GWs, we set $\eta = {\eta}_0$ to obtain the WKB solution at present, where $a_0 = 1$. 

The modes that become sub prior to the AD era, experience the whole AD era, and thus their amplitudes behave as $|\chi_k|={e^{-d}}a_{\times}$. Conversely, the $k$ modes that become sub after the AD era when $A$ is negligible, do not experience any suppression from the $d$ factor and they go as $|\chi_k|= a_{\times}$. If a $k$ mode becomes sub during the AD era, the suppression factor on its amplitude will depend on its $k$ value: $|\chi_k|={e^{-d_k}}a_{\times}$.

Finally, let us comment on the behavior of $a_{\times}$ as a function of $k$ to gain an intuition about $|\chi_k|$ in different $k$ regimes. Since for the cosmological eras with constant $\omega$, $a \propto \eta^{\frac{2}{1 + 3 w}}$, we can obtain $a_{\times}$ given the functional form of $\eta_{\times}$ in terms of $k$: $k \sim a_{\times}H_{\times} = \frac{1}{\eta_{\times}}$, leading to  $a_\times \propto k^{\frac{-2}{1+3w}}$. However, this argument is \textit{only} valid when the effect of $A$ is ignored (before and after AD era when $H \gg A$). 
For the modes that enter the horizon during $AD$, this simple analysis does not suffice and numerical analysis is required. 

\subsection{The Power Spectrum}\label{subsecPS}
To introduce the power spectrum of gravitational waves as an observable, we need their energy density given by \cite{lino2022gravitational, saikawa2018primordial}
\begin{align}\label{2-3-38}
\rho_{GW} = \frac{1}{32\pi G} \langle \dot{h}_{ij} (t, x) \dot{h}_{ij} (t, x) \rangle_{av}
= \frac{1}{32\pi G} \sum_{\zeta = +, \times} \int \frac{\text{d}^3 k}{(2\pi)^3} 2 \lvert \dot{h}_k^{(\zeta)} \rvert^2,
\end{align}
where $\langle ... \rangle_{av}$ denotes spatial averaging. The power spectrum, $\Omega_{GW}(t,k)$, is defined as the energy density of GWs per logarithmic frequency interval, divided by the critical energy density, $\rho_{\text{crit}} = 3H^2(8\pi G)^{-1}$, of the Universe \cite{lino2022gravitational, saikawa2018primordial, d2019imprint}: $\Omega_{GW} (t, k) = \text{d} \rho_{GW} (t, k)/(\rho_{\text{crit}}\text{d}\ln k)$. Using the parametrization $h_k (\eta) = h^{\text{Prim}}_k \chi_k(\eta)$, we obtain \cite{lino2022gravitational, saikawa2018primordial}
\begin{equation}\label{2.62}
\Omega_{GW} (\eta, k) = \frac{1}{12 a^2 (\eta) H^2 (\eta)} \mathcal{P}_T(k) \left(\chi^\prime(\eta, k)\right)^2.
\end{equation} 
\begin{figure}
	\centering
	\includegraphics[width=0.65\textwidth]{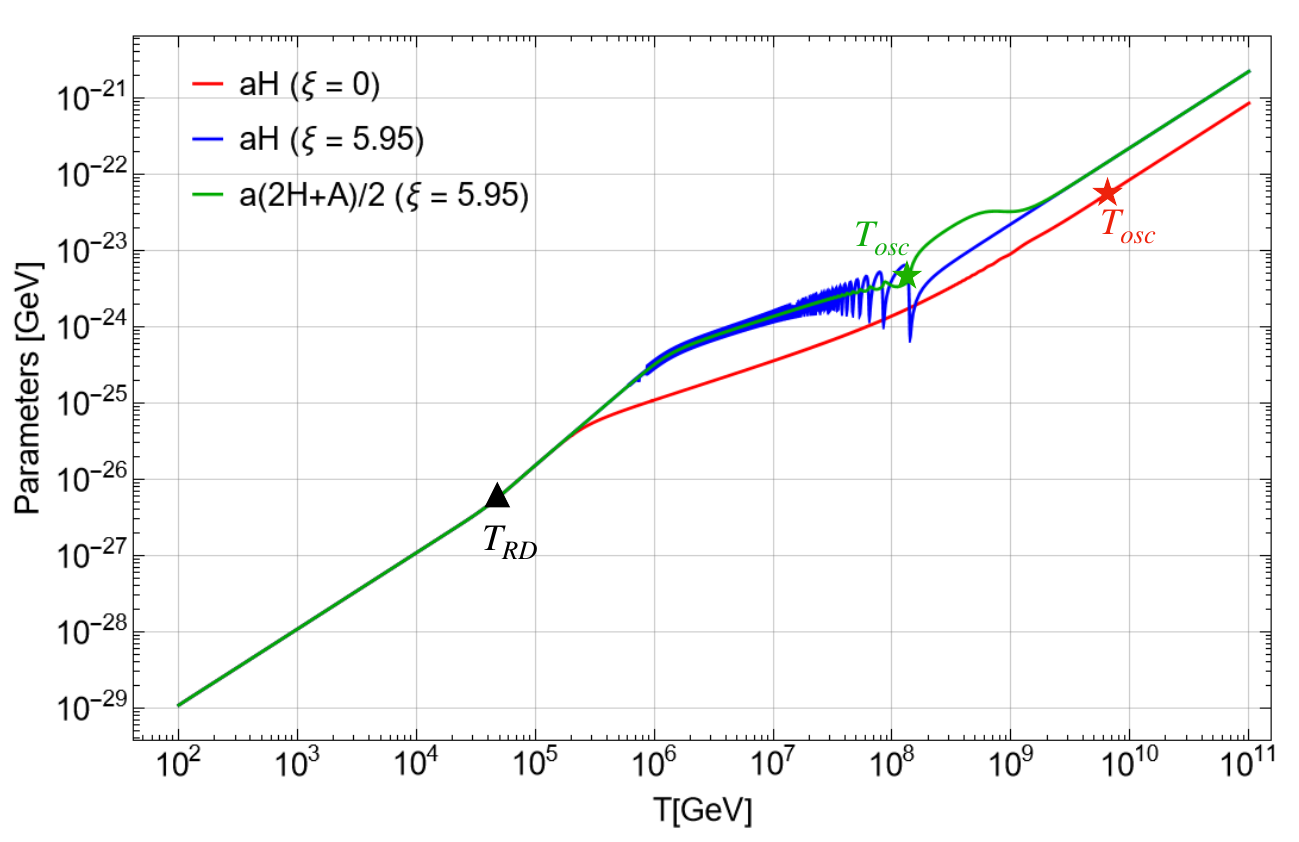}
	\caption{This plot demonstrates the evolution of  $\frac{a}{2}(2H+A)$ with respect to temperature.  The red line is for the case of $\xi = 0$, the blue line is for $\xi = 5.95$ while ignoring the $A$ term, and the green line presents the case of $\xi = 5.95$ keeping the $A$ term. Notice that since the $A$ term is zero (ignored) in the red (blue) line, we have  $\frac{a}{2}(2H+A) \to a H$. In this plot, $T_{osc}$s are shown by stars,  and $T_{RD}$ is represented by a triangle. As the definition of $T_{osc}$ does not depend on $A$, the temperature of oscillation for the blue and green lines coincide. Furthermore, $T_{RD}$ is independent of the value of $\xi$, and thus all of the lines in the plot share the same $T_{RD}$.}\label{aHfig}
\end{figure}
The quantity $\mathcal{P}_T(k)$ is the primordial tensor power spectrum which is a nearly scale-invariant power-law function of $k$ around a pivot scale $k_*$ as $\mathcal{P}_T (k) = A_T \left(k/k_*\right)^{n_T}$. Here, $A_T$ is the amplitude and $n_T$ is the tilt of the spectrum \cite{maggiore2018gravitational, gorbunov2011introduction, d2019imprint}. Using the WKB solution, Eq.~(\ref{WKB}) for the sub modes $k \gg \frac{a}{2}(2H+A)$, we get $|\chi^\prime (\eta, k)| \simeq \ k| \chi (\eta,k)|$ (see Appendix\ \ref{WKBAp}). By substituting this derivative in Eq.\ (\ref{2.62}) and setting $\eta= \eta_0$, the present power spectrum for GWs is obtained as 
\begin{equation}\label{2.63}
\Omega_{GW}^0 (f) \simeq \frac{1}{12} \frac{k^2}{a_0^2 H_0^2} \mathcal{P}_T(k) \lvert \chi_k \rvert^2,
\end{equation}
where $f = k/(2\pi a_0)$, $\chi_k$, $a_0$, and $H_0 = 100h\ km/s/Mpc$ are the present time quantities. In this work, we set $A_T \simeq 1.5 \times 10^{-10}$, $n_T = 0.4$ and $k_\star = k_{\text{CMB}} = 0.05~\frac{1}{Mpc}$ in $\mathcal{P}_T(k)$ as is done in \cite{d2019imprint} to access the maximal reach of future detectors. These values are consistent with the upper bound on the tensor-to-scalar ratio $r = \frac{A_T}{A_R} \le 0.07$, where $A_R \simeq 2.1\times 10^{-9}$ is the amplitude of scalar perturbations obtained from observational data \cite{akrami2020planck}.\footnote{Both scalar and tensor perturbations have scale-invariant power spectrums, and their properties are predicted by the inflationary models. However, the properties of the former are well studied through observations \cite{gorbunov2011introduction}.} Eq.(\ref{2.63}) elegantly shows how the parametrization $h_k (\eta) = h^{\text{Prim}}_k \chi_k(\eta)$ shows up in the power spectrum. Therefore, $\Omega_{GW}^0$ is generally made by concerning two factors: 1) \textit{the initial condition $h^{Prim}_k$} and 2) \textit{the value of $\lvert \chi_k \rvert^2$} (influenced by the indirect and direct effects).
Note that GWs start from after becoming sub, coded in the primordial tensor power spectrum, ${\mathcal{P}_T}(k)$. 
The form of power spectrum $\Omega_{GW}^{0}$ for the high-frequency modes (becoming sub before AD) and low-frequency ones (becoming sub after AD) using $|\chi_k|$ obtained in Sec.~\ref{EGW} would be:
\begin{align}
\Omega_{GW}^{0(High)} (k) &\propto e^{-2d} k^n,\label{2-3-44}\\   
\Omega_{GW}^{0(Low)} (k) &\propto k^n,\label{2-3-55}
\end{align}
where $n = n_T + \frac{2(3\omega - 1)}{3 \omega + 1}$. In the case of $\xi =0$, or equivalently $A = 0$, the present power spectrum is roughly $\Omega_{GW}^{0} (k) \propto k^n$ \cite{d2019imprint}. 

\begin{figure}
	\centering
	\includegraphics[width=0.65\textwidth]{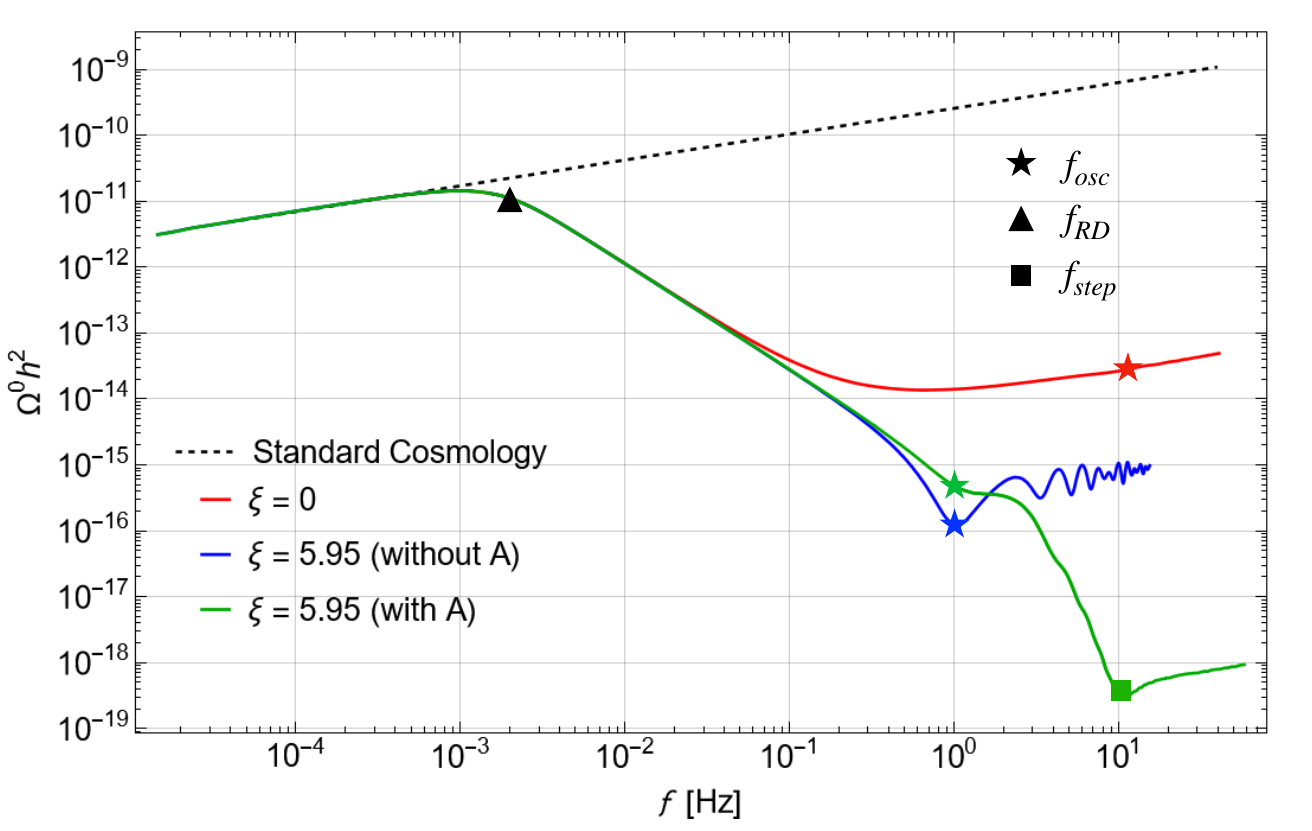}
	\caption{The power spectrum of GWs for standard cosmology (dashed black line), $\xi = 0$ (red line), $\xi = 5.95$ ignoring the $A$ term (blue line), and $\xi = 5.95$ preserving the $A$ term (green line) with respect to frequency is presented. The frequencies corresponding to the beginning of $\phi$ oscillation (Eq.~(\ref{eq:fosc})), late RD era (Eq.~(\ref{eq:fRD})), and the step (Eq.~(\ref{eq:fstep})) are shown by star, triangle, and square, respectively. The ripples of the blue line are due to the intense oscillation of $\phi$ which is felt by the modes that become sub during the first RD era. In the case of the green line, the presence of $A$ damps the ripples. Moreover, the $A$ term leads to a noticeable drop in the power spectrum of the green line at high frequencies, manifesting a step-like feature.}\label{fig:10}
\end{figure}

\subsection{Frequencies $f_{osc}$ and $f_{RD}$}\label{sec33}
Before diving into the numerical results on power spectrum, let us discuss the relation between a $k$ mode, or more precisely its corresponding frequency $f = k/(2\pi a_0)$, and the temperature at which it becomes sub. Since this happens when $k \simeq\frac{a}{2}(2H+A)$, we can study Fig.~\ref{aHfig} to find the transition frequencies. In this figure, the red line represents the $\xi =0$ case, the green line is for $\xi=5.95$, and we have included the blue line where the case of $\xi=5.95$ is redrawn neglecting the effect of $A$. It is important to emphasize the blue line is not a physical case; however, it can illustrate the role of $A$ more lucidly. 
Other than this numerical method using Fig.~\ref{aHfig}, we can use the following theoretical expression to find the frequency at $T = T_{RD}$ \cite{d2019imprint}:
\begin{equation}
f_{RD} \simeq b f_0 \left(\frac{g^{s,0}_{*}}{g^{s}_{*RD}}\right)^{\frac{1}{3}} \left(\frac{g^{\rho}_{*RD}}{g^{\rho, 0}_{*}}\right)^{\frac{1}{2}}  \left(\frac{\Omega^{0}_{R}}{1/2}\right)^{\frac{1}{2}} \frac{T_{RD}}{T_0},
\label{eq:fRD}
\end{equation}
where $T_0 \simeq 2.72548 K \simeq 2.3 \times 10^{-13}$GeV is the CMB temperature \cite{Fixsen:2009ug}, $h^2\Omega^0_R \simeq 4.2\times10^{-5}$ is the present energy density parameter of radiation, and $f_0 = \frac{H_0}{2\pi}$ is the frequency of the GW mode, with comoving wavelength of one Hubble radius, $f_0/h\simeq5.2\times10^{-19}$Hz. The parameter $b$ is obtained numerically as $b\simeq 1.4$, in our study. Furthermore, the frequency $f_{osc}$ corresponding to the mode that becomes sub at $T_{osc}$ can be obtained     
using Eqs.~(\ref{eq:Tosc}) and (\ref{eq:TRD}):
\begin{equation}
f_{osc} \simeq f_{RD} \left(\frac{m_{\phi}c_{RD}}{i_{osc}\alpha_{RD}\Gamma} \right)^{1-\alpha_{RD}} \left[\frac{1+\xi\left(\frac{\phi_{in}}{M_{Pl}}\right)^2}{1+\xi(6\xi-1)\left(\frac{\phi_{in}}{M_{Pl}}\right)^2}\right]^{\frac{1-\alpha_{RD}}{2}}.
\label{eq:fosc}
\end{equation}
%where $f_{RD}$ is given by Eq.\ (\ref{eq:fRD}).          
Having these frequencies we can determine our scenario how narrates on power spectrum $\Omega_{GW}^0(f)$.

\begin{figure}[t!]
	\centering
	\includegraphics[width=0.65\textwidth]{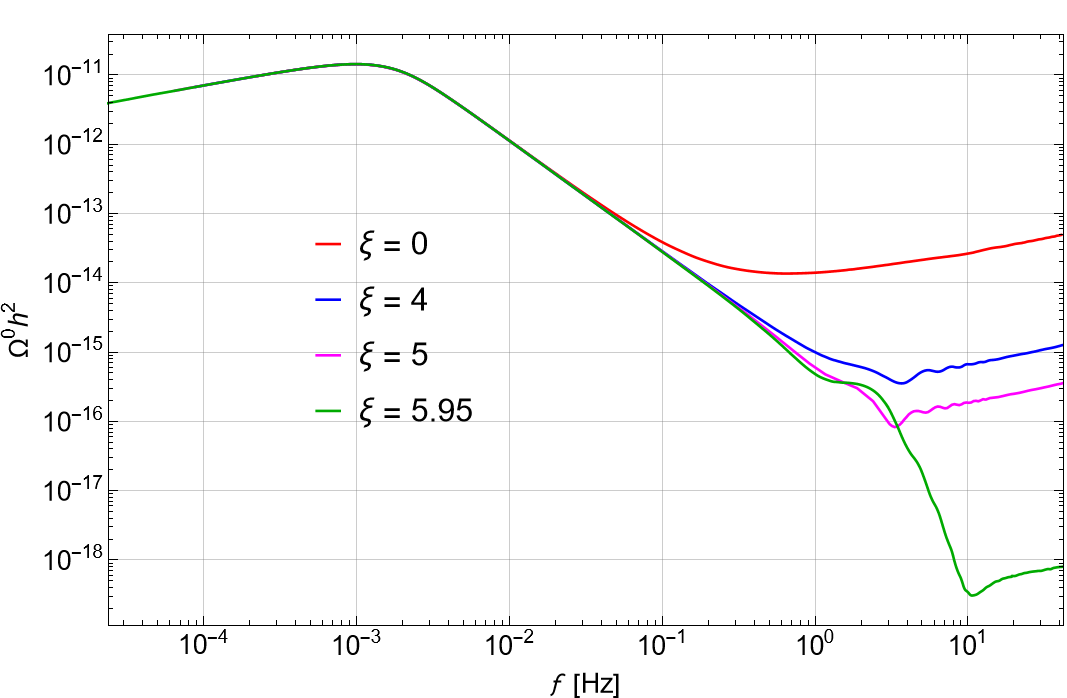}
	\caption{ The power spectrum for different values of scalar-gravity coupling $\xi$ is shown. Increasing $\xi$
		results in a decrease in the power spectrum at high frequencies. Since both the Hubble rate and $A$ are proportional to $ \left(\frac{1}{8 \pi G } - \xi \phi^2\right)^{-1}$, increasing $\xi$ such that $\xi \phi^2 \simeq \frac{1}{8 \pi G }$ significantly increases the Hubble rate and $A$. Therefore, the drop at high frequencies for the case of $\xi = 5.95$ (green line) is considerably bigger than $\xi = 5$ (magenta line) or $\xi = 4$ (blue line).}\label{fig:7}
\end{figure}

\section{Results}\label{results}

To obtain the power spectrum $\Omega_{GW}^0$, we need to evolve GWs till today. However, given that after $\phi$ decay, we return to the standard cosmology, we can use simple scaling arguments to determine $\Omega_{GW}^0$ from the power spectrum after $\phi$'s effect has been diminished. To do this, we solve Eq.~(\ref{transeq}) from $T_{in} = 10^{11}~\text{GeV}$ to $T_{f} = 100~\text{GeV}$, then scale $|\chi_k(T_f)|$ to its present value $|\chi_k|$ using the conservation of entropy, $S(T)\propto g_{*}^s(T) a^3 T^3 $ \cite{saikawa2018primordial}.
In this section, our main goal is to investigate how the non-minimal ($\xi \geq 1$) gravitational coupling of the scalar field, $\xi R \phi^2$, affects the power spectrum. In particular, we want to distinguish the direct and indirect effects on $\Omega_{GW}^0$. This splitting can be done since they show up independently in the WKB solution, Eq.~(\ref{WKB}). To proceed, we study the evolution of $\chi_k$ with and without the $A$ term for $\xi= 5.95$, and contrast them with the case of $\xi = 0$. \footnote{We can only study the modes that become sub after $T_{in}$ and oscillate for a sufficient time before $T_f$ ($a_{f} H_{f} \ll k \ll a_{in} H_{in}$). 
Hence, the accessible frequency interval in this work is $10^{-5}~\text{Hz}\ll f \ll10^{2}~\text{Hz}$.}

Fig.~\ref{fig:10} demonstrates the values of $\Omega_{GW}^0$ for the case of standard cosmology (dashed black line), the case of $\xi = 0$ (red line), and the non-minimal gravity coupling case $\xi=5.95$, with (green line) and without (blue line) the $A$ term. The frequencies that become sub during the RD era exhibit an almost scale-invariant spectrum. 
Introducing a transient MD era, causes high-frequency modes to be lower relative to the case of standard cosmology. That is because during the MD era, the damping term in Eq.~(\ref{transeq}) becomes $\frac{a^\prime}{a} = aH \propto \frac{2}{\eta}$, which is greater than that of the RD era, $\frac{a^\prime}{a} = aH \propto \frac{1}{\eta}$. Modes that become sub during the MD era, experience a weaker damping.  

The aforementioned effects on $\Omega_{GW}^0$ due to the MD era are generally the same for both $\xi=0$ and $\xi>0$. However, the presence of the $\xi R \phi^2$ term leaves a distinct imprint on the shape of the power spectrum. To better comprehend these effects, let us compare the blue line ($\xi = 5.95$, while ignoring the A term) in Fig.~\ref{fig:10} with the red line ($\xi = 0$). The main difference between these two lines is the change in the Hubble rate due to the $\xi R \phi^2$ term. The main points to highlight from the comparison of the blue and red lines are: 1) the oscillations of $H$ appear in the power spectrum; 2) the MD era becomes longer\footnote{The reason MD era becomes longer with the increase in $\xi$ is because $R < 0$, or equivalently $m_{eff}< m_{\phi}$ at high temperatures. Since the start of oscillations is roughly when $H \simeq m_{eff}$, the Hubble rate at the beginning of oscillation is lower for the case of non-minimal coupling, which means that $\phi$ gets less damped and consequently the MD era gets longer.} making a deeper kink in $\Omega_{GW}^0$; 3) the temperature at which the Late RD era begins, coincides with that of the case $\xi = 0$.

To grasp the effect of the $A$ term on $\Omega_{GW}^0$, we compare the green line ($\xi=5.95$, while preserving the $A$ term) with the blue line ($\xi=5.95$, ignoring the $A$ term) in  Fig.~\ref{fig:10}. As can be seen from these two lines, the $A$ term has a notable effect on the shape of the power spectrum. One effect is that the oscillations on $\Omega_{GW}^0$ vanish, since the oscillations of $aA$ cancel those of $2aH$. Furthermore, an additional step-like feature appears in $\Omega_{GW}^0$, which can be understood by the extra damping term in Eq.~(\ref{transeq}) during the AD era. We can find the temperature at which this step occurs, $T_{step}$ by setting  $A \simeq 2H/\beta$, where the parameter $\beta$ depends on $\xi$. Using this along with the form of A in terms of $\phi$ and Eq.~(\ref{eq:Hubble}), the exact expression for $\dot \phi^2$ at $T_{step}$ is obtained as:
\begin{equation}
\dot \phi^2=\frac{m^2_{\phi } {\phi}^2_{in}+2\rho_R}{6\beta(\beta+2)\xi^2\gamma-1},
\label{eq:phidot2}
\end{equation}
where $\gamma=\frac{\phi^2_{in}}{M^2_{Pl}-\xi\phi^2_{in}}$. Substituting $A$ obtained from Eq.~(\ref{eq:phidot2}) and $H\simeq \sqrt{\frac{\rho_R }{3 (M_{Pl}^2- \xi \phi^2_{in})}}$ in $A \simeq 2H/\beta$, we obtain
\begin{align}
T_{step} &\simeq \left(\frac{90 \beta^2 \xi^2 \gamma m^2_{\phi } {\phi}^2_{in}}{\pi^2 g^{\rho}_{* step}(12\beta\xi^2\gamma-1)}\right)^{\frac{1}{4}},\label{eq:Tstep}\\
f_{step} &\simeq f_{osc}\left[\frac{\beta^2 \xi^2 \gamma^2 i_{osc}\left[1+\xi(6\xi-1)\left(\frac{\phi_{in}}{M_{Pl}}\right)^2\right]}{(12\beta\xi^2\gamma-1)\left[1+\xi\left(\frac{\phi_{in}}{M_{Pl}}\right)^2\right]}\right]^{\frac{1-\alpha_{step}}{2}}\left(1+\frac{1}{\beta}\right),
\label{eq:fstep}
\end{align}
where $\alpha_{step} \simeq 0.57$. This frequency is shown by a square in Fig.~\ref{fig:10}. 
To complete this part of our study, we obtain $\Omega_{GW}^0$ for different values of $\xi$, and present the results in Fig.~\ref{fig:7}. As can be seen, the kink in $\Omega_{GW}^0$ deepens by increasing $\xi$, as expected.
\begin{table}[t!]
	\caption{This table presents the values of (1) $\alpha_{_{RD, D, step}}$ and $\beta$ obtained from numerical matching, (2) $f_{_{RD, osc, step}}$ according to theoretical expectation and according to numerical analysis, (3) the dilution factor (4) the damping factor $e^{-2d}$, and (5) the ratio of $\Omega^0_{GW}$ for each of the mentioned frequencies according to both theory and data analysis, for different values of $\xi$. Albeit one needs some numerical input for the theoretical prediction, this table shows that Eqs. ( \ref{eq:omegatoomega}), (\ref{eq:omegatoomega2}), and (\ref{eq:omegatoomega3}) match the numerical results with satisfactory accuracy.   }
	\label{table3}
	\begin{center}\scriptsize{
			\hspace*{-1.7cm}\begin{tabular}{|p{2.5cm}||c|c|c|c|c|c|c|c|c|}
				\hline
				${\xi} $			&3&4&4.5&5&5.5&5.6&5.7&5.9&5.95   \\ \hline\hline
				$ \alpha_{_{RD}}$	&0.637808&0.640833&0.642508&0.643449&0.644169&0.643961&0.644907&0.644607&0.644885   \\ \hline
				$ \alpha_{_{D}}$	&0.66138&0.672266&0.677025&0.681421&0.685498&0.686401&0.686925&0.688673&0.688931  \\ \hline
				$ \alpha_{_{step}}$	&0.474328&0.537462&0.574286&0.599133&0.60678&0.61608&0.608802&0.595607&0.567124   \\ \hline
				$ \beta $			&81.8583&48.311&48.6144&40.0857&24.5725&21.2776&17.5458&20.8901&20.9099   \\ \hline
				$ f_{_{RD}}|_{_{theory}} $	&0.001647&0.001651&0.001653&0.001654&0.001655&0.001655&0.001656&0.001656&0.001656	  \\ \hline
				$ f_{_{RD}}|_{_{data}} $	&0.001633&0.001633&0.001634&0.001634&0.001634&0.001628&0.001634&0.001628&0.001633	  \\ \hline
				$ f_{_{osc}}|_{_{theory}} $	&1.351&1.182&1.110&1.061&1.022&1.020&0.999&0.9950&0.9880	  \\ \hline
				$ f_{_{osc}}|_{_{data}} $ &1.338&1.169&1.097&1.048&1.008&1.003&0.9888&0.9780&0.9741	  \\ \hline
				$ f_{_{step}}|_{_{theory}} $ &4.437&3.661&3.488&3.388&3.503&3.510&3.664&5.409&10.85	 \\ \hline
				$ f_{_{step}}|_{_{data}} $	&4.397&3.621&3.447&3.347&3.457&3.451&3.628&5.316&10.70	  \\ \hline
				
				$ Dilution\; Factor$	&0.05052&0.03879&0.03359&0.02835&0.02219&0.02057&0.01880&0.01242&0.006301   \\ \hline
				$ \exp(-2d)$	&0.4992&0.3320&0.2467&0.1624&0.07793&0.06084&0.0439&0.009229&0.0006203   \\\hline
				$  \frac{\Omega_{_{osc}}}{\Omega_{_{RD}}}^\star|_{_{theory}}\times 10^{6} $ &96.07&47.81&35.25&26.77&20.78&19.68&18.98&17.07&16.80   \\ \hline
				
				$  \frac{\Omega_{_{osc}}}{\Omega_{_{RD}}}^\star|_{_{data}}\times 10^{6} $ &102.8&44.19&30.1&20.31&13.93&12.76&11.98&10.16&9.874   \\ \hline
				
				$ \frac{\Omega_{_{step}}}{\Omega_{_{osc}}}|_{_{theory}}$ &1.025&0.371&0.1803&0.08505&0.03558&0.02410&0.01894&0.003925&0.0003890  \\ \hline	
				
				$ \frac{\Omega_{_{step}}}{\Omega_{_{osc}}}|_{_{data}} $ &0.7453&0.4151&0.268&0.1503&0.06200&0.04814&0.03591&0.007590&0.0005911  \\ \hline	
				
				$ \frac{\Omega_{_{step}}}{\Omega_{_{RD}}}|_{_{theory}}\times 10^{6} $ &79.08&26.00&14.32&7.267&2.871&2.15&1.564&0.3406&0.02983   \\ \hline
				
				$ \frac{\Omega_{_{step}}}{\Omega_{_{RD}}}|_{_{data}}\times 10^{6} $ &95.64&27.93&14.26&6.524&2.359&1.799&1.345&0.2846&0.02341   \\ \hline			
		\end{tabular}}
	\end{center}
	%	\scriptsize{\raggedright  $ \frac{\Omega_{_{Osc}}}{\Omega_{_{RD}}}^\star  $ obtained from {\color{Green} the case} without A. \par}
\end{table}

Another way to interpret our results is by using \textit{dilution factor}: $D \equiv \left[S(T \gg T_{osc}) / S(T \ll T_{RD})\right]^{1/3}$, where $S=a^3s$ with $s=\frac{2\pi^2}{45}g_{*}^sT^3$ being the comoving entropy density. Using $a \propto t^\alpha \simeq (\frac{\alpha}{H})^\alpha$, and $H_{osc} \simeq \sqrt{\frac{\rho^{osc}_R}{3(M^2_{Pl}-\xi \phi^2_{in})}}$ and $H_{RD} = \sqrt{\frac{2\rho^{RD}_R}{3M^2_{Pl}}}$, along with $T_{osc}$ and $T_{RD}$ given by Eqs.~(\ref{eq:Tosc}) and (\ref{eq:TRD}), the dilution factor becomes
\begin{equation}
D \simeq (\frac{m_\phi c_{RD}}{i_{osc}\alpha_{RD}\Gamma})^{\frac{1-2\alpha_D}{2}}  \left(\frac{g^{s}_{* osc}}{g^{s}_{*RD}}\right)^{\frac{1}{3}} \left(\frac{2g^{\rho}_{*RD}}{g^{\rho}_{*osc}}\right)^{\frac{1}{4}}\left[1-\xi\left(\frac{\phi_{in}}{M_{Pl}}\right)^2\right]^{\frac{1}{4}} \left[\frac{1+\xi\left(\frac{\phi_{in}}{M_{Pl}}\right)^2}{1+\xi(6\xi-1)\left(\frac{\phi_{in}}{M_{Pl}}\right)^2}\right]^{\frac{1-2\alpha_D}{4}},
\label{eq:DF2}
\end{equation}
with $\alpha_D \approx 0.68$ obtained from numerical matching. To interpret the results of Fig.~\ref{fig:7}, we employ the WKB solution (see Appendix \ref{WKBAp}). For the sub mode with the wave number $k$, we have $\lvert \chi_k \rvert \simeq a_k$. The parameter $a_k$ is the scale factor at which the $k$ mode becomes sub. Replacing $\lvert \chi_k \rvert$ in Eq.~(\ref{2.63}) with $a_k=k/H (a_k)$, for the modes with frequencies $f_{osc}$ and $f_{RD}$, we obtain $\Omega^0_{GW}(f_{osc})/\Omega^0_{GW}(f_{RD})$ (ignoring $A$ term) as:
\begin{align}\label{eq:omegatoomega}
\begin{split}
\frac{\Omega^0_{GW}(f_{osc})}{\Omega^0_{GW}(f_{RD})} \simeq \left(\frac{f_{osc}}{f_{RD}}\right)^{n_T}D^4\left(\frac{g^{s}_{* RD}}{g^{s}_{*osc}}\right)^{\frac{4}{3}}\left(\frac{g^{\rho
}_{*osc}}{2g^{\rho}_{*RD}}\right)\left[1-\xi\left(\frac{\phi_{in}}{M_{Pl}}\right)^2\right]^{-1},%\\
\end{split}
\end{align}
\begin{figure}
	\centering
	\includegraphics[width=0.65\textwidth]{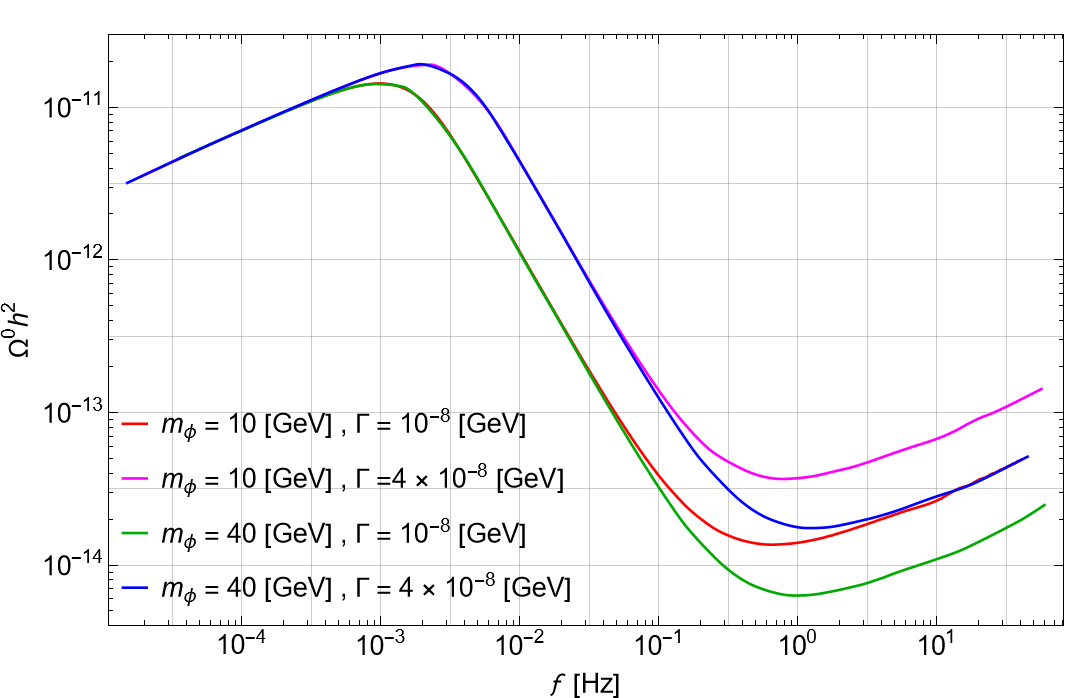}
	\caption{ The power spectrum for four cases: (1) unchanged mass and decay rate with values as in Fig.~\ref{fig:10} (red line), (2) increased mass (green line), (3) increased decay rate (magenta line), (4) increased mass and the decay rate but unchanged $m/\Gamma$ value (blue line).}\label{fig:5}
\end{figure}
which demonstrates the indirect effect of the term $\xi R \phi^2$. Second, if we set the wave number as $k = k_{step}$, we have $\lvert \chi_k \rvert \simeq e^{-d} a_k$. Replacing $\lvert \chi_k \rvert$ in Eq.\ (\ref{2.63}) with $\frac{e^{-d}k}{H (a_k)(1+\frac{1}{\beta})}$, for this mode, while using $\frac{H_{osc}}{H_{step}}\simeq \left[\frac{f_{osc}}{f_{step}}(1+\frac{1}{\beta})\right]^{\frac{1}{1-\alpha_{step}}}$, we obtain
\begin{equation}
\frac{\Omega^0_{GW}(f_{step})}{\Omega^0_{GW}(f_{osc})}\simeq e^{-2d} \left(\frac{f_{step}}{f_{osc}}\right)^{n_T+4-\frac{2}{1-\alpha_{step}}}\left(1+\frac{1}{\beta}\right)^{\frac{2\alpha_{step}}{1-\alpha_{step}}},
\label{eq:omegatoomega2}
\end{equation}
where $d = \frac{1}{2} \int^{\frac{10^5}{m_{\phi}}}_{t_{step}}A\text{d}t $. Eq.\ (\ref{eq:omegatoomega2}) exhibits the direct effect of the term $\xi R \phi^2$. Finally, implementing the correction parameter $cr$ as the ratio of $h^2 \Omega^0_{GW}(f_{osc})$ in the power spectrum with $A$ to that of the power spectrum without $A$, we can write
\begin{equation}
\frac{\Omega^0_{GW}(f_{step})}{\Omega^0_{GW}(f_{RD})}\simeq cr \frac{\Omega^0_{GW}(f_{step})}{\Omega^0_{GW}(f_{osc})} \frac{\Omega^0_{GW}(f_{osc})}{\Omega^0_{GW}(f_{RD})},
\label{eq:omegatoomega3}
\end{equation}
where $\frac{\Omega^0_{GW}(f_{step})}{\Omega^0_{GW}(f_{osc})}$ and $\frac{\Omega^0_{GW}(f_{osc})}{\Omega^0_{GW}(f_{RD})}$ are given by Eqs.~(\ref{eq:omegatoomega2}) and (\ref{eq:omegatoomega}), respectively. To test our estimations, we compare some analytical results with numerical ones. Table~\ref{table3} compares the analytical estimates and data driven pivotal frequencies $f_{_{RD, osc, step}}$ for different values of $\xi$. Furthermore, the ratio of $\Omega^0_{GW}$ at these key frequencies according to the analytical expectations (Eqs. (\ref{eq:omegatoomega}), (\ref{eq:omegatoomega2}), and (\ref{eq:omegatoomega3})), and the data are provided in Table~\ref{table3} as well. As it is illustrated, the predicted values are consistent with the numerical data.  Albeit the analytical expressions depend on data driven parameters $\alpha_{_{RD, D, step}}$ and $\beta$, they prove to be valuable as they show the dependence of our observable $\Omega^0_{GW}(f)$ on model parameters and initial conditions. 

\begin{figure}
	\centering
	\includegraphics[width=0.8\textwidth]{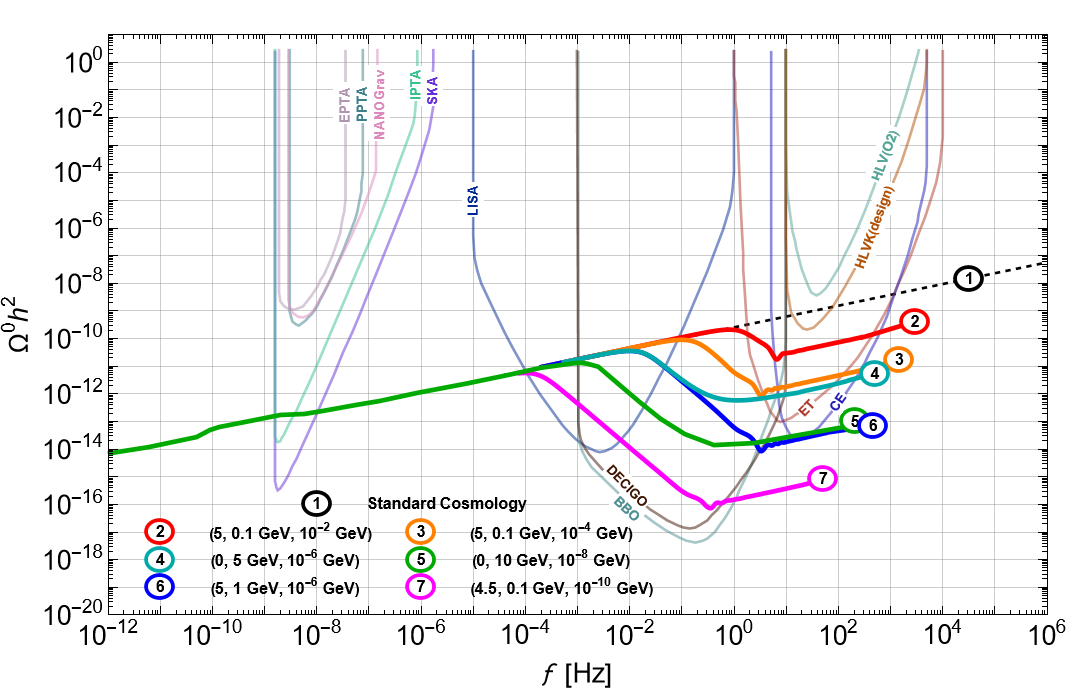}
	\caption{The sensitivity curves of existing and future GW experiments and the power spectrum of GWs in $\xi=0$ and $\xi\neq0$ scenarios. There is hope that the effect of direct and indirect effects of modified gravity $\xi R \phi^2$ can be captured for specific regions in the parameter space of the model by future detectors. The curves are resulted using different benchmarks of ($\xi,\;m_{\phi},\;\Gamma$). Since we only can analyze the modes that become sub \textit{after} $T_{in}=10^{11}$ GeV, the colored solid lines that are the results of the paper do not extend to arbitrarily large frequencies.}\label{fig:8}
\end{figure}

For the sake of completeness, let us study the effect of $m_\phi$ and $\Gamma$ on the shape of $\Omega_{GW}^0$, in the case of $\xi=0$. Here we choose $m_{\phi}=10\ \text{GeV},\ 40\ \text{GeV}$ and $\Gamma=10^{-8}\ \text{ GeV}, \ 4 \times 10^{-8} \ \text{GeV} $. Since $m_\phi$ determines the beginning of the MD era and $\Gamma$ controls when it ends, only the position of the kinks change and the slopes in $\Omega_{GW}^0$ stay constant, as shown in Fig.\ \ref{fig:5}. 
Finally, the effects of non-minimal gravitational coupling, $\xi R \phi^2$, of the scalar field on $\Omega_{GW}^0$ can be probed by future GWs experiments. Due to the shape of $\Omega_{GW}^0$, we expect GWs to be detectable by some of the observatories focusing on low-frequencies but missed by high-frequency experiments. The shape of the GWs may also be probed by some experiments that can detect relatively lower-intensity GWs.   In Fig. \ref{fig:8}, we have picked some benchmarks that may be probed by LISA \cite{LISA:2017pwj}, BBO \cite{Harry:2006fi}, and DECIGO \cite{Kawamura:2011zz}.

%%%%%%%%%%%%%%%%%%%%%%%%%%%%%%%%%%%%%%%%%%%%%%%%%%%%%%%%%%%
\section{Conclusion}\label{con}

In this paper, we investigated the evolution of the SGWB originating from inflation within a cosmological framework. We considered a scenario where a scalar field $\phi$ dominates the energy density of the Universe at high temperatures and has a non-minimal coupling with gravity, represented as $\xi R \phi^2$. Previous studies have demonstrated that early matter domination leads to a reduced signal strength of SGWB in the present. Various experiments such as LISA, BBO, and DECIGO can probe SGWB, while others like HLVK and ET may miss it. 

By introducing a non-minimal coupling between $\phi$ and gravity, not only does this decrease amplify, but non-trivial features also emerge at certain frequencies. When the coupling parameter $\xi$ is large, the damping term in the evolution of SGWB is enhanced due to a larger Hubble rate and the presence of a $\xi$-dependent term denoted as $A$ in this study. The distinctive step-like feature observed in Fig.\ \ref{fig:8} is caused by the dominance of $A$ over the Hubble rate. Analytically, this feature can be understood by examining the ratio $\frac{\Omega^0_{GW}(f_{step})}{\Omega^0_{GW}(f_{RD})}$, which includes not only the dilution factor but also an additional damping term $e^{-\int A dt}$ compared to the case where $\xi = 0$. Additionally, when $\xi \neq 0$, the Hubble rate is relatively higher at the beginning of the MD era, resulting in a lower value of $\Omega_{GW} h^2$, even if the $A$ term is neglected, as the dilution factor discloses. We provided an analytical expression for $\Omega^0_{GW}$ at key frequencies and demonstrated that they align with the numerical results with acceptable precision.

In the scenario of early matter domination with $\xi = 0$, the power spectrum pattern exhibits deviations from a pure power law at frequencies determined by the mass and decay width of $\phi$. Our study reveals that introducing the term $\xi R \phi^2$ does not significantly alter the location of these deviations but introduces a distinct step feature that serves as a signature of this term in observational data. We identified a set of benchmarks that exhibit this feature within the parameter space where DECIGO and BBO experiments can probe. In general, the shape of the power spectrum of Primordial Gravitational Waves (PGWs) as a function of frequency provides valuable insights into the evolution of the early universe.

\section*{Acknowledgment}

We thank Enrico Morgante, Nicklas Ramberg, Wolfram Ratzinger,  Nematollah Riazi, Kai Schmitz, and Pedro Schwaller for useful discussions. F.E. and H.M. are grateful to CERN for their hospitality. The work in Mainz is supported by the Cluster of Excellence Precision Physics, Fundamental Interactions and Structure of Matter (PRISMA$^+$ – EXC 2118/1) within the German Excellence Strategy (project ID 39083149). Also, F.E. is supported by the grant 05H18UMCA1 of the German Federal Ministry for Education and Research (BMBF), and H.M. is supported by the National Natural Science Foundation of China under Grants No. 12247107 and 12075007.    

%%%%%%%%%%%%%%%%%%%%%%%%%%%%%%%%%%%%%%%%%%%%%%%%%%%%%%%%%%%

\appendix 

\section{ Beginning of the evolution: Attractor Behavior, Hubble parameter, Ricci scalar, and EoS parameter}\label{appa}
\label{app:attractor}
This appendix shows that the benchmarks satisfying Eq.\ (\ref{eq:inicond}) have an attractor behavior. That is, even if we start with $\dot \phi_{in}  \neq 0 $, we will quickly merge to the path of $\phi = \phi_{in}$ and $\dot \phi_{in} = 0$ in the phase space. Our other objective in this appendix is to show that even in the case of non-minimal coupling where $\xi \geq 1$, the start of evolution is effectively RD  defined as $\omega = \frac{1}{3}$, even though the energy density of the scalar field dominates over that of radiation $\rho_\phi > \rho_R$.

For better understanding, we rewrite the EoM of $\phi$, Eq.(\ref{2.6}), with the assumption $H \gg \Gamma$ at the beginning as
\begin{equation}
\ddot \phi + 3 H \dot \phi + m^2_{eff} \phi =0.
\end{equation}
 This equation has the form of a harmonic oscillator equation, with $m_{eff}$ as the angular frequency and the friction term of $\frac{3H}{2}$. In the limit where $\frac{3H}{2} \gg m_{eff}$, and with the initial conditions $\phi(T_{in})=\phi_{in}$ and $\dot{\phi}(T_{in}) = 0$ the solution for $\phi$ becomes $\phi = \phi_{in} = const$, which means that the field is stuck at its initial value. If the initial velocity of the harmonic oscillator is non-zero, the solution to the equation in this regime becomes 
\begin{equation}
\phi(t) = \phi_{in} + \frac{ \dot \phi_{in}}{3 H} - \frac{\dot \phi}{3H}e^{- 3 H t}, 
\label{eq:gensol1}
\end{equation}
which shows that in this case, the general solution does not have constant behavior and in fact is time-dependent. For any value of $\dot{\phi}$ satisfying Eq.\ (\ref{eq:inicond}), the second term in Eq.\ (\ref{eq:gensol1}) becomes negligible compared with $\phi_{in}$ and hence we can write the solution as $\phi(t) = \phi_{in} - \frac{\dot \phi}{3H}e^{- 3 H t}$. In addition, the third term rapidly approaches zero with time, and hence $\phi$ approaches $\phi_{in}$ again at high temperatures which obviously demonstrates the attractor behavior mentioned above.

To obtain a general understanding of the evolution of the background at the beginning, and besides that an estimation for the depth of the kink in the GWs power spectrum, it is essential to find an approximate relation for the Hubble parameter. Starting from the first Friedmann equation, Eq.\ (\ref{2.15}) and using the energy density of scalar field with non-minimal gravitational coupling, Eq.\ (\ref{eq:rhophi}), we have: 
\begin{equation}
H = \sqrt{\frac{1}{ 3  M_{Pl}^{2} }(\rho_R + \rho_\phi)}= \frac{ \xi  \phi \dot \phi + \sqrt{ \xi ^2 \phi^2 \dot \phi^2+ \frac{1}{3} \left(M_{Pl}^2 -\xi  \phi^2\right) \left(\frac{m_\phi ^2 \phi^2}{2}+ \frac{\dot \phi^2}{2} +\rho_R\right)}}{ {M_{Pl}^2- \xi  \phi^2}}.
\label{eq:Hubble}
\end{equation}
We are considering a benchmark class that satisfies Eq.\ (\ref{eq:inicond}). Therefore, we can simplify the Hubble rate at $T_{in}$ to the following: 
\begin{align}
H (T \simeq T_{in}) & \simeq \frac{ \xi  \phi_{in} \dot \phi_{in} + \sqrt{ \xi ^2 \phi^2_{in} \dot \phi^2_{in}+ \frac{1}{3} \left(M_{Pl}^2 -\xi  \phi^2_{in}\right) \rho_R (T_{in})}}{M_{Pl}^2- \xi  \phi^2_{in}}\nonumber \\
& \simeq \sqrt{\frac{\rho_R (T_{in})}{3 (M_{Pl}^2- \xi  \phi^2_{in})}},
\label{eq:Happrox}
\end{align}
where in the first line $m_\phi^2 \phi_{in}^2/ 2, \dot \phi^2_{in}/2 \ll \rho_R (T_{in}) $, and in the second line $ \frac{ 3 \xi^2 \dot \phi_{in}^2 \phi_{in}^2}{M_{Pl}^2 - \xi \phi_{in}^2 } \ll \rho_{R} (T_{in})$ is used. This equation demonstrates that the initial Hubble parameter increases by increasing $\xi$, which can be seen for the two values of $\xi = 0$ and $\xi  = 5.95$ in Fig\ \ref{fig:1}. As long as these conditions are satisfied, Eq.\ (\ref{eq:Happrox}) provides the approximate Hubble rate even if we move away from $T_{in}$.
With the Hubble parameter in hand, the EoS parameter at the beginning of the evolution can be found using the relation \cite{opferkuch2019ricci, dimopoulos2018non}
\begin{equation}\label{omeg}
R=3(1-3\omega)H^2.
\end{equation}
Hence, to have an estimation of the $\omega$ at the beginning of the evolution, we need the behavior of $R$ at high temperatures. To this end, we take the following steps. First, by contracting Eq.\ (\ref{E2}) with the metric, the trace of the equation yields 
\begin{equation}\label{eq:Riccigen}
R=\frac{-1}{M_{Pl}^2}g^{\mu\nu}(T_{\mu\nu}^{(M)}+T_{\mu\nu}^{(\phi)})=-\frac{1}{M_{Pl}^2}(T^{(M)}+T^{(\phi)}).
\end{equation} 
To use the above equation, we need to derive the trace of $T_{\mu\nu}^{(\phi)}$ by contacting with metric \cite{figueroa2021dynamics}
\begin{equation}\label{2-13}
T^{(\phi)}=(6\xi-1)(\partial^\mu \phi \partial_\mu \phi + \xi R \phi^2)+6\xi,
\end{equation}
Now, substituting Eq.~(\ref{2-13}) in Eq.~(\ref{eq:Riccigen}), gives the Ricci scalar as \cite{figueroa2021dynamics}:
\begin{equation}\label{2-15}
R=\frac{(1-6\xi)\partial^\mu \phi \partial_\mu \phi+4 \left(\frac{1}{2}m_\phi^2 \phi^2\right)-6\xi m_{\phi^2}^2 \phi^2-T^{(M)}}{M_{Pl}^2+(6\xi-1)\xi\phi^2}.
\end{equation}
In our study, the scalar field $\phi$ is assumed to be homogeneous and isotropic (and hence, just a function of time). On the other hand, $T^{(M)}=0$, since in our study, the matter part is only the radiation, and the trace of the energy-momentum tensor for relativistic fluid with $\omega = \frac{1}{3}$ is zero. As a result, the Ricci scalar at the beginning becomes
\begin{equation}
R (T \simeq T_{in}) = \frac{(1- 6 \xi) \dot \phi_{in}^2+ (2- 6 \xi) m_{\phi}^2\phi_{in}^2}{M_{Pl}^2 + (6 \xi -1) \xi \phi_{in}^2}.
\label{eq:Riccifull}
\end{equation}
According to Eq.\ (\ref{omeg}), in order to evaluate $\omega$, we need to find the dimensionless parameter $R/H^2$:
\begin{equation}
\frac{R}{H^2}(T \simeq T_{in}) \simeq \frac{(1- 6 \xi) \dot \phi_{in}^2+ (2- 6 \xi) m_{\phi}^2\phi_{in}^2}{M_{Pl}^2 + (6 \xi -1) \xi \phi_{in}^2} \times \frac{3 (M_{Pl}^2- \xi  \phi^2_{in})}{\rho_R (T_{in})} \to  0, 
\label{eq:RoverH2}
\end{equation}
because we are assuming $\rho_R (T_{in}) \gg \dot \phi_{in}, m_{\phi}^2\phi_{in}^2$. Now, given Eq.\ (\ref{eq:RoverH2}), we see that $\omega \simeq 1/3$ for the class of all benchmarks satisfying Eq.\ (\ref{eq:inicond}).

\section{The WKB Analysis}\label{WKBAp}
In this appendix, we explain the WKB analysis for the evolution of the modes in sub-regime in the presence of the term, A\cite{odintsov2022spectrum, nishizawa2018generalized, arai2018generalized}. Such an additional term appears in the various modified gravity theories and hence, our analysis in this appendix is quite general and we do not restrict ourselves to the case of $\xi R \phi^2$. This analysis provides us with physical intuition and parametrizes the solution in terms of the aforementioned direct and indirect effects on the evolution of GWs. Starting from Eq. (\ref{2-3-47}), it is conventional to rewrite this equation as:
\begin{equation}\label{2.69}
\chi^{\prime\prime} + \mathcal{H} (2 + M) \chi^\prime + k^2 \chi = 0,
\end{equation}
where we invoke the parametrization $h (\eta) = h^{\text{Prim}} \chi (\eta)$, $\mathcal{H}= \frac{a'}{a}$ and $M = \frac{A}{H}$. To obtain the WKB solution for high-frequency modes in this case, we consider an ansatz: $\chi= Z e^{iY}$ where $C=Z h^{prim}$. Substituting this ansatz for the transfer function in Eq.\ (\ref{2.69}), separates the equation into two equations for the imaginary and the real part:
\begin{align}
k^{2} + (2+M) \mathcal{H}& \frac{Z'}{Z} - (Y')^{2} +\frac{Z''}{Z}=0,\label{2.72}\\ 
(2+M)\mathcal{H}&+2\frac{Z'}{Z}+\frac{Z''}{Z'}=0.\label{2.73}
\end{align}
Since our aim here is to derive the WKB solution for high-frequency modes, we can neglect the second and fourth terms in Eq.\ (\ref{2.72}). That is because high-frequency modes correspond to the modes that enter the horizon at early times, and thus the terms involving $Z$ are negligible compared to the terms that include the phase of $\chi$. Hence, from Eq. (\ref{2.72}), we have:
\begin{equation}
Y=\pm k\eta.
\end{equation}
Substituting this result in Eq. (\ref{2.73}), we simply have the WKB solution \cite{odintsov2022spectrum, nishizawa2018generalized, arai2018generalized}:
\begin{align}
\begin{split}
\chi &\propto e^{-\int\left(1+\frac{M}{2}\right)\mathcal{H}\text{d}\eta} e^{\pm ik \eta}\\
& = e^{-\int \mathcal{H}\text{d}\eta} \times e^{\pm ik \eta} \times e^{-\frac{1}{2} \int M \mathcal{H}\text{d}\eta}  
\end{split}
\end{align}
Concentrating on this solution, we see that the first two exponentials can be simplified to $
e^{-\int\mathcal{H}\text{d}\eta} e^{\pm ik \eta} = e^{-\int\mathcal{H}\text{d}t} e^{\pm ik \eta}= \frac{1}{a} e^{\pm ik \eta}$
if we rewrite it in cosmic time. This part of the WKB solution is exactly the form of the WKB solution in the standard GR (cases of the standard cosmology and the case of $\xi = 0$) and hence, conventionally called $\chi_{GR}$.\footnote{This terminology may be confusing. In the case of modified gravity, the evolution of background is modified, which through $a$ and $H$, affects the evolution of GWs (indirect effects). Hence, the evolution of $a$ in the $\chi_{GR}$ is modified and carries all the effects of the modified evolution of the background, in contrast with the standard GR. This terminology only refers to the fact that the functional form of this part of the solution, is like the WKB solution in the standard GR.} As we can see $\chi_{GR}$ demonstrates the oscillatory behavior of the high-frequency modes in the sub-regime. It is conventional to write the complete WKB solution for $\xi \neq 0 $ as:
\begin{equation}
\chi = C_{Mod} \chi_{GR},
\end{equation}
where
\begin{equation}
C_{Mod} \equiv e^{-d},\qquad d\equiv  \frac{1}{2} \int M \mathcal{H}\text{d}\eta,
\end{equation}
and $\chi_{GR} = \frac{1}{a} e^{\pm ik\eta}$. The effect of the additional damping term in the equation of GWs appears in the additional exponential $e^{-d}$. The parameter d can be written in cosmic time as:
\begin{equation}
d = \frac{1}{2} \int M \mathcal{H}\text{d}\eta = \frac{1}{2} \int M H\text{d}\eta = \frac{1}{2} \int A\text{d}t
\end{equation}
As before, all we need to know about the GWs are well coded in the observable of GWs, the power spectrum, introduced in Eq.\ (\ref{2.62}). It will be more convenient if we try to write it in terms of the transfer function itself. It is in fact possible for sub-modes where we have $k \gg \frac{a}{2}\left(2H + A\right)$, and $A > 0$ (in our specific case). Starting from
\begin{equation}
\chi = e^{-\frac{1}{2} \int_{\eta_{\times}}^{\eta} aA \,d\eta} \chi_{GR},
\end{equation}
and invoking the useful relation
\begin{equation}
\frac{d}{dx}  \int_{g(x)}^{f(x)} h(t) \,dt = h(f(x))f'(x)-h(g(x))g'(x),
\end{equation}
we can simply conclude that
\begin{equation}
|\chi'| \simeq k|\chi|.  
\end{equation}
By substituting the above equation in Eq.\ (\ref{2.62}), Eq.\ (\ref{2.63}) is obtained.

\bibliography{GW.bib}
\bibliographystyle{JHEP}
\end{document}